\documentclass[twocolumn]{aastex63}
\usepackage{CJK}
\usepackage{amsmath}
\usepackage{verbatim}

\usepackage{hyperref}
\usepackage{multirow}

\shorttitle{A Gemini/GMOS Transmission Spectrum of WD~1856+534~b}
\shortauthors{Xu et al.}

\begin{document}
\begin{CJK}{UTF8}{gbsn}
\title{Gemini/GMOS Transmission Spectroscopy of the Grazing Planet Candidate WD~1856+534~b}

\correspondingauthor{Siyi Xu} 
\email{siyi.xu@noirlab.edu}

\author[0000-0002-8808-4282]{Siyi Xu (许\CJKfamily{bsmi}偲\CJKfamily{gbsn}艺)}
\affil{Gemini Observatory/NSF's NOIRLab, 670 N. A'ohoku Place, Hilo, Hawaii, 96720, USA}

\author[0000-0001-8274-6639]{Hannah Diamond-Lowe}
\thanks{These two authors contributed equally to this manuscript.}
\affil{National Space Institute, Technical University of Denmark, Elektrovej, 2800 Kgs.\ Lyngby, Denmark}

\author[0000-0003-4816-3469]{Ryan J. MacDonald}
\thanks{These two authors contributed equally to this manuscript.}
\affil{Department of Astronomy and Carl Sagan Institute, Cornell University, 122 Sciences Drive, Ithaca, NY 14853, USA}

\author[0000-0001-7246-5438]{Andrew Vanderburg}
\affil{Department of Physics and Kavli Institute for Astrophysics and Space Research, Massachusetts Institute of Technology, 77 Massachusetts Avenue, Cambridge, MA 02139, USA}
\affil{Department of Astronomy, University of Wisconsin-Madison, Madison, WI, USA}

\author[0000-0002-9632-1436]{Simon Blouin}
\affil{Los Alamos National Laboratory, P.O. Box 1663, Mail Stop P365, Los Alamos, NM 87545, USA}

\author[0000-0003-4609-4500]{P. Dufour}
\affil{Institut de Recherche sur les Exoplan\`etes (iREx), Universit\'e de Montr\'eal, Montr´eal, QC H3C 3J7, Canada}
\affil{D´epartement de physique, Universit\'e de Montr\'eal, Montr\'eal, QC H3C 3J7, Canada}

\author[0000-0002-8518-9601]{Peter Gao}
\thanks{NHFP Sagan Fellow}
\affil{Department of Astronomy and Astrophysics, University of California Santa Cruz, Santa Cruz, CA 95064, USA}

\author[0000-0003-0514-1147]{Laura Kreidberg}
\affil{Max-Planck-Institut f\"{u}r Astronomie, K\"{o}nigstuhl 17, 69117 Heidelberg, Germany}

\author[0000-0002-3681-2989]{S. K. Leggett}
\affil{Gemini Observatory/NSF's NOIRLab, 670 N. A'ohoku Place, Hilo, Hawaii, 96720, USA}

\author[0000-0002-4128-6901]{Andrew W. Mann}
\affil{Department of Physics and Astronomy, The University of North Carolina at Chapel Hill, Chapel Hill, NC 27599, USA}

\author[0000-0002-4404-0456]{Caroline V. Morley}
\affil{Department of Astronomy, University of Texas at Austin, Austin, TX, USA}

\author[0000-0002-4434-2307]{Andrew W. Stephens}
\affil{Gemini Observatory/NSF's NOIRLab, 670 N. A'ohoku Place, Hilo, Hawaii, 96720, USA}

\author[0000-0003-3987-3776]{Christopher E. O'Connor}
\affil{Department of Astronomy and Carl Sagan Institute, Cornell University, 122 Sciences Drive, Ithaca, NY 14853, USA}

\author[0000-0001-5729-6576]{Pa Chia Thao}
\thanks{NSF GRFP}
\affil{Department of Physics and Astronomy, The University of North Carolina at Chapel Hill, Chapel Hill, NC 27599, USA}

\author[0000-0002-8507-1304]{Nikole K. Lewis}
\affil{Department of Astronomy and Carl Sagan Institute, Cornell University, 122 Sciences Drive, Ithaca, NY 14853, USA}

\begin{abstract}

WD~1856+534~b is a Jupiter-sized, cool giant planet candidate transiting the white dwarf \object{WD~1856+534}. Here, we report an optical transmission spectrum of WD~1856+534~b obtained from ten transits using the Gemini Multi-Object Spectrograph. This system is challenging to observe due to the faintness of the host star and the short transit duration. Nevertheless, our phase-folded white light curve reached a precision of 0.12 \%. WD~1856+534~b provides a unique transit configuration compared to other known exoplanets: the planet is $8\times$ larger than its star and occults over half of the stellar disc during mid-transit. Consequently, many standard modeling assumptions do not hold. We introduce the concept of a `limb darkening corrected, time-averaged transmission spectrum' and propose that this is more suitable than $(R_{\mathrm{p}, \lambda} / R_{\mathrm{s}})^2$ for comparisons to atmospheric models for planets with grazing transits. We also present a modified radiative transfer prescription. Though the transmission spectrum shows no prominent absorption features, it is sufficiently precise to constrain the mass of WD~1856+534~b to be $>$ 0.84 M$_\mathrm{J}$ (to $2 \, \sigma$ confidence), assuming a clear atmosphere and a Jovian composition. High-altitude cloud decks can allow lower masses. WD~1856+534~b could have formed either as a result of common envelope evolution or migration under the Kozai-Lidov mechanism. Further studies of WD~1856+534~b, alongside new dedicated searches for substellar objects around white dwarfs, will shed further light on the mysteries of post-main sequence planetary systems.

\end{abstract}

\keywords{White dwarf stars -- Exoplanet atmospheres -- Extrasolar gaseous planets -- Brown Dwarfs }

\section{Introduction}

Planetary systems are ubiquitous around main-sequence stars, but little is known about their fate once the host star leaves the main sequence. There is indirect evidence that planetary systems are present and active around white dwarfs. For example, up to half of all white dwarfs display atmospheric pollution, which is interpreted as a result of accretion of extrasolar planetesimals \citep{Zuckerman2003,Zuckerman2010,Koester2014a}. The most heavily polluted white dwarfs often display an infrared excess from a circumstellar dust disk \citep[e.g.][]{Farihi2009,Chen2020} and occasionally double peaked emission features from a circumstellar gas disk \citep[e.g.][]{Gaensicke2006,Dennihy2020b}. Dynamical perturbation and tidal disruption of extrasolar planetesimals is the most widely accepted explanation for these systems \citep{Jura2003}. Different mechanisms, such as planet-planet scattering, the Kozai-Lidov effect, and galactic tides can bring an initially far away planetesimal close to the tidal radius of the white dwarf \citep[e.g.][]{DebesSigurdsson2002,BonsorVeras2015, Veras2016,Stephan2017}. Subsequently, the disrupted planetesimal may be circularized under the white dwarf's radiation or through the dust and gas drag and form a debris disk, which resides within the white dwarf's tidal radius \citep{Veras2014I,Veras2015II,Malamud2021}.

Recent surveys have enabled many discoveries of planetary systems orbiting white dwarfs. The extended {\it Kepler} mission found WD~1145+017 displays deep transit features every 4.5~hours \citep{Vanderburg2015}. Followup observations confirmed that the transits are variable on a daily basis, consistent with an asteroid in the process of actively disintegrating around the white dwarf \citep[e.g.][]{Rappaport2016,Gary2017}. The Zwicky Transient Facility (ZTF) survey has detected a few more white dwarfs with similar transit signatures \citep{Vanderbosch2020,Guidry2020, vanderbosch2021}. In addition, \citet{Manser2019} reported a stable 123~min variation in the calcium infrared triplet emission around the white dwarf SDSS~J1228+1043, which is interpreted to come from an iron-rich object in close orbit. The Sloan Digital Sky Survey (SDSS) spectra of WD~J0914+1914 show many exotic double-peaked emission features. Initially, the system was thought to be an interacting binary, but later it was recognized that the most likely scenario is evaporation or tidal disruption of a Neptune-like planet \citep{Gaensicke2019}. These recent discoveries suggest that intact planets must also be present around white dwarfs, but until recently there was no direct detection of such objects.

The first planet candidate transiting a white dwarf, WD~1856+534~b, was discovered orbiting \object{WD~1856+534} using data from the Transiting Exoplanet Survey Satellite ({\it TESS}; \citealt{Vanderburg2020}). The transit signal is 8~min long, 56\% deep, and repeats every 1.4~days. Transit models show that WD~1856+534~b is consistent with an object roughly the size of Jupiter at a distance of 0.02~AU from the white dwarf. The non-detection of thermal emission from the planet in the {\it Spitzer} band implies a temperature upper limit of 290~K and a mass upper limit of 13.8 Jupiter mass (M$_\mathrm{J}$, to $2 \, \sigma$ confidence, \citealt{Vanderburg2020}). Unfortunately, radial velocity observations to constrain the planetary mass are not possible due to the lack of spectral features in the white dwarf's spectrum. Recently, \citet{Alonso2021} obtained a transmission spectrum of WD~1856+534~b with the Gran Telescopio Canarias (GTC) OSIRIS and EMIR instruments in an attempt to constrain WD~1856+534~b's mass. The flat transmission spectrum provides an upper limit on the scale height of the planet's atmosphere, and hence a lower limit on the mass of 2.4~M$_\mathrm{J}$ at 2$\sigma$ \citep{Alonso2021}. In addition, \object{WD~1856+534} is in a triple star system with a binary M dwarf pair G~229-20~A/B about 40{\farcs}0 away. The discovery of WD~1856+534~b has triggered significant interest in its origin: how did it end up so close to its white dwarf?

In this paper, we present a transmission spectrum of WD~1856+534~b spanning ten transit epochs using the Gemini Multi-Object Spectrograph (GMOS) at the Gemini Telescope. We describe the observing strategy and data reduction in Section~\ref{sec:obs} and present updated parameters of \object{WD~1856+534} in Section~\ref{sec:wd}. In Section~\ref{sec:LCs}, we introduce a new concept---a limb darkening corrected, time-averaged transmission spectrum---for modeling grazing transit transmission spectra and apply it to WD~1856+534~b. We describe a modified radiative transfer prescription for grazing transits and present our results, including a revised mass limit for WD~1856+534~b, in Section~\ref{sec:theory_models_WD1856b}. Finally, we discuss the implications of this study in Section~\ref{sec:dis} and summarize our conclusions in Section~\ref{sec:con}.

\section{Observation \& Data Reduction \label{sec:obs}}

\subsection{Observation}

We used the GMOS instrument \citep{Hook2004} at the Gemini North Telescope to observe ten transits of WD~1856+534~b under programs GN-2020A-DD-108 and GN-2020B-Q-131 (PI: S. Xu). The observing strategy follows previous GMOS transmission spectroscopy observations \citep{Gibson2013a, Gibson2013b, Stevenson2014, Huitson2017}. The multi-object spectroscopy (MOS) mode was used, which allowed for simultaneous observations of the target \object{WD~1856+534} and the reference star TIC~267574908. The basic information of these two objects is listed in Table~\ref{tab:target}. We chose the reference star based on its similar brightness and effective temperature to WD~1856+534, as well as proximity. The MOS science mask is designed with two slits centered on the target and the reference star, respectively. Each slit is 10{\farcs}0 wide in the dispersion direction (to minimize slit losses) and 30{\farcs}0 long in the spacial direction (to sample the sky background). We also used the R150 grating with a central wavelength of 0.69~$\mathrm{\mu}$m, which has a spectral resolution of 631 for a 0{\farcs}5 wide slit. The actual spectral resolution of the observation varied depending on the seeing condition. In addition, the OG515 filter was used to block second order contamination. In order to wavelength calibrate our data, we also designed a MOS calibration mask identical to the science mask but with slit widths of 1{\farcs}0. 

\begin{deluxetable}{lcccccccccc}
\tablecaption{ \label{tab:target} Basic information about the target and the reference star}
\tablehead{
 & \colhead{Target} & \colhead{Reference Star } 
}
\startdata
Name & WD 1856+534 & TIC~267574908\\
RA\tablenotemark{$\dag$}  &18:57:39.8 & 18:57:41.2\\
Dec\tablenotemark{$\dag$} &+53:30:32.5 & +53:31:30.5\\
G (mag)\tablenotemark{$\dag$} & 17.0& 15.9\\
Spectral type & DA  & G \\
T$_\mathrm{eff}$ (K) & 4860 & 5600\tablenotemark{$\ddag$}\\
  \enddata
\tablenotetext{$\dag$}{\cite{GaiaDR2}.}
\tablenotetext{$\ddag$}{\cite{Stassun2019}.}
\end{deluxetable}

Each observing block is about 40~min long, centered at the transit of \object{WD~1856+534}, which lasts a total of 8~min \citep{Vanderburg2020}. The individual exposure time is 60~sec. To increase the duty cycle, we used 2 by 2 binning and a custom region of interest (ROI) for readout. The ROI is defined as two rectangles centered on the target and the reference star, respectively, each being 274~pixels wide in the spatial direction and 3072~pixels long in the spectral direction. The average readout time is 25~sec, giving a duty cycle of 70\%. In addition, three on-sky flats were taken immediately before and after each transit observation to minimize the impact of flexure. Three arcs using the calibration mask were also taken after the observations for wavelength calibration. The observing log is provided in Table~\ref{tab:log}. A total of ten transits were observed under different weather conditions.

\begin{deluxetable*}{clccccccccccc}
\tablecaption{ \label{tab:log} Observing Log}
\tablehead{
\colhead{\#} & \colhead{Date} & \colhead{Exposure }  & \colhead{Number of} & \colhead{Air Mass} & \multicolumn{3}{c}{Observing Conditions\tablenotemark{a}} \\
 & \colhead{(UTC)} & Time (s)  &Exposures  & & \colhead{Image Quality}& \colhead{Cloud Coverage} & \colhead{Sky Background} 
}
\startdata
1	& 2020 Jun 15, 09:35:55--10:09:01 & 60  & 23 & 1.27--1.34 & 85\% & 50\% & 20\% \\
2	& 2020 Jul 02, 07:04:27--07:43:37 & 60 & 27 & 1.47--1.64 & 70\%& 50\%& 100\%\\
3	& 2020 Jul 30, 10:42:22--11:20:42 & 60 & 27 & 1.31--1.41 & 20\%& 50\%& 80\%\\
4	& 2020 Aug 23, 09:00:59--09:45:15\tablenotemark{b} & 60 & 27 & 1.30--1.40 & 85\%& 50\%& 20\%\\
5	& 2020 Aug 30, 10:02:08--10:46:40 & 60 & 33 & 1.57--1.82 & 70\%& 50\%&100\%\\
6	& 2020 Sep 02, 05:40:41--06:18:23 & 60 & 27 & 1.20--1.22 & 20\%& 80\%& 100\%\\
7	& 2020 Sep 09, 06:38:46--07:16:33 & 60 & 26 & 1.21--1.24 & 20\%& 50\%& 20\%\\
8	& 2020 Sep 23, 08:15:16--09:15:40\tablenotemark{c} & 60 & 32 & 1.52--1.85 & 70\%& 50\%& 50\%\\
9	& 2020 Oct 03, 05:05:12--05:44:15 & 60 & 27 & 1.21--1.24 & 20\%& 50\%& 80\%\\
10	& 2020 Oct 17, 06:56:21--07:39:51 & 60 & 30 & 1.59--1.84 & 70\%& 50\%& 50\%\\
  \enddata
\tablenotetext{a}{The observing conditions are defined in terms of percentile. Image Quality 20\%is about 0{\farcs}5 measured on the science frames and 85\%is about 1{\farcs}5. Cloud coverage 50\% is considered photometric conditions with no flux loss, while 80\% has about 1~mag of extinction. Sky background 20\% has a sky brightness $\mu_\mathrm{V}$ $>$ 21.3~mag, and 100\% has $\mu_\mathrm{V}$ $>$ 18~mag. See details in  \href{https://www.gemini.edu/observing/telescopes-and-sites/sites\#Constraints}{www.gemini.edu/observing/telescopes-and-sites/sites\#Constraints}.}
\tablenotetext{b}{ This observation was performed unguided due to the mediocre seeing condition.}
\tablenotetext{c}{ There is a gap between 08:58:04--09:08:51 due to an observing error (the calibrations were executed instead of science observations).}
\end{deluxetable*}

\subsection{Data Reduction}
We used the Gemini IRAF package\footnote{\href{https://www.gemini.edu/observing/phase-iii/understanding-and-processing-data/getting-started\#gmos}{www.gemini.edu/observing/phase-iii/understanding-and-processing-data/getting-started\#gmos}} for bias subtraction, flat correction, cosmic ray rejection, wavelength calibration, and spectral extraction. We adjusted some parameters, since the combination of the MOS mode and the custom ROI readout is a rare GMOS setup and not officially supported by the data reduction package. With the new Hamamatsu CCD \citep{Scharwaechter2018}, the slight misalignment between the slit mask and the CCD detector described in previous GMOS transmission spectroscopy is no longer an issue \citep{Stevenson2014}. Therefore, we adopted a standard sky subtraction routine. At each epoch, we extracted the spectra using an aperture radius of three times the Full Width at Half Maximum (FWHM) measured in the science frames. We propagated the uncertainty in each reduction step by turning on the fl\_vardq flag. 

\begin{figure*}
\epsscale{1.15}
\plotone{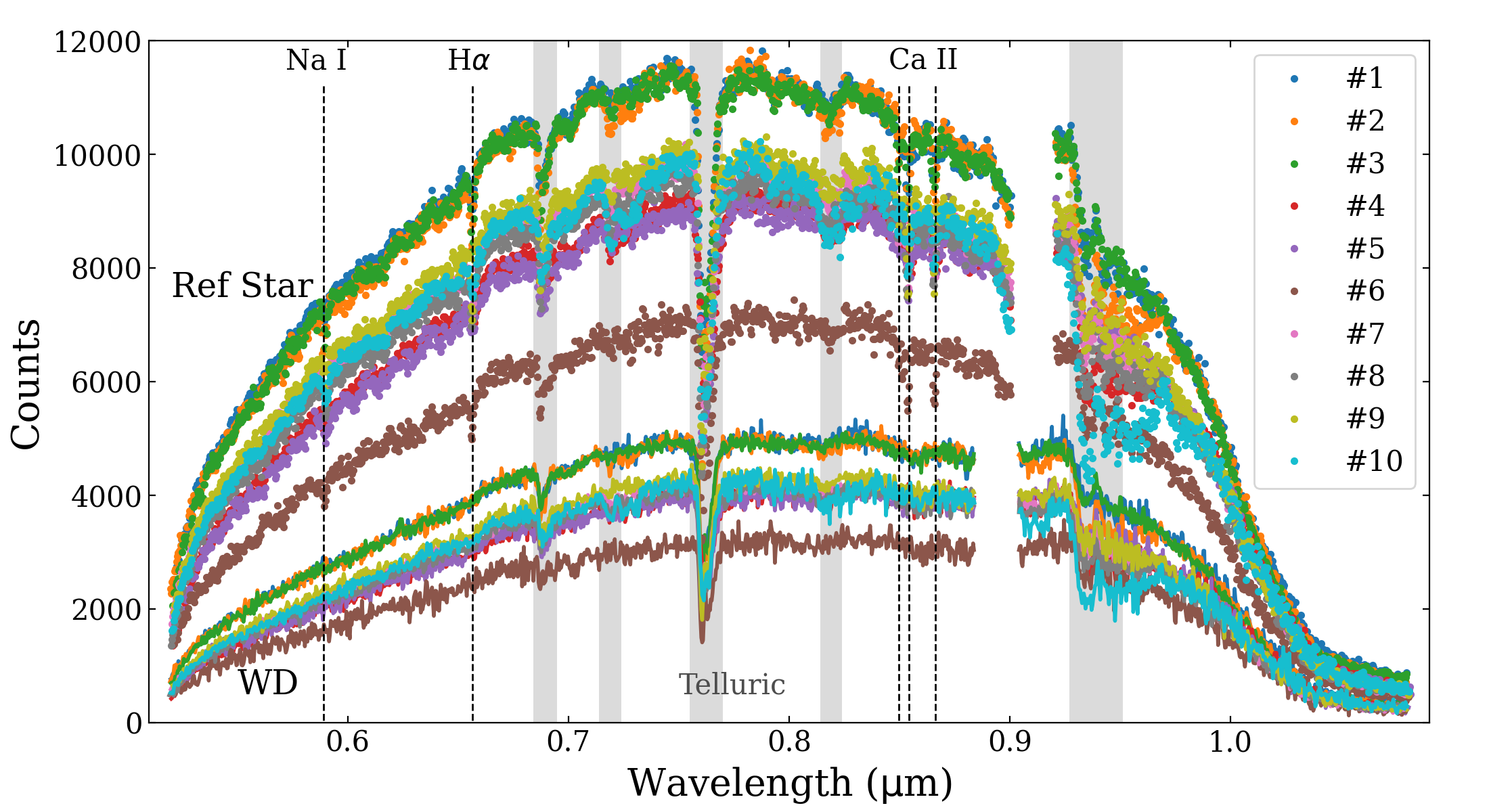}
\caption{Representative spectra of the white dwarf and the reference star. Different color represents different epochs listed in Table~\ref{tab:log}. The dotted and solid lines represent the reference star and the white dwarf, respectively. Telluric features are highlighted in grey. In the reference star, atmospheric absorption from H$\alpha$, Ca~II and Na~I are detected. 
}
\label{fig:spectra}
\end{figure*}

Figure~\ref{fig:spectra} shows representative spectra of the target and the reference star. The wavelength coverage is 0.52--1.08~$\mathrm{\mu}$m, with a small gap due to the GMOS CCD chip gap. Most of the observed features in the white dwarf's spectra are from the Earth's atmosphere. In the combined out-of-transit spectra, there is weak absorption from H$\alpha$ (see Section~\ref{sec:wd}). For the reference star, we detect some absorption features from the star's atmosphere in addition to the telluric features. To correct for possible wavelength shifts between different epochs, we cross-correlate the atmospheric O$_\mathrm{2}$ A band around 0.76~$\mathrm{\mu}$m in both the white dwarf spectra and the reference star spectra with a telluric spectrum. As shown in Figure~\ref{fig:spectra}, the average counts in transits \#4--10 are about 20\% lower than those in transits \#1--3 due to a telescope throughput issue caused by the primary mirror. Transit \#6 has the lowest counts because of high cloud coverage (see Table~\ref{tab:log}). Otherwise, the data quality in different epochs is comparable. Transit \#4 was performed unguided, which, interestingly, had little impact on the quality of the data.

\section{White Dwarf Parameters \label{sec:wd}}

\begin{figure*}
\centering
\includegraphics[width=0.495\textwidth, trim={0.5cm 0.0cm 0.0cm 0.0cm}]{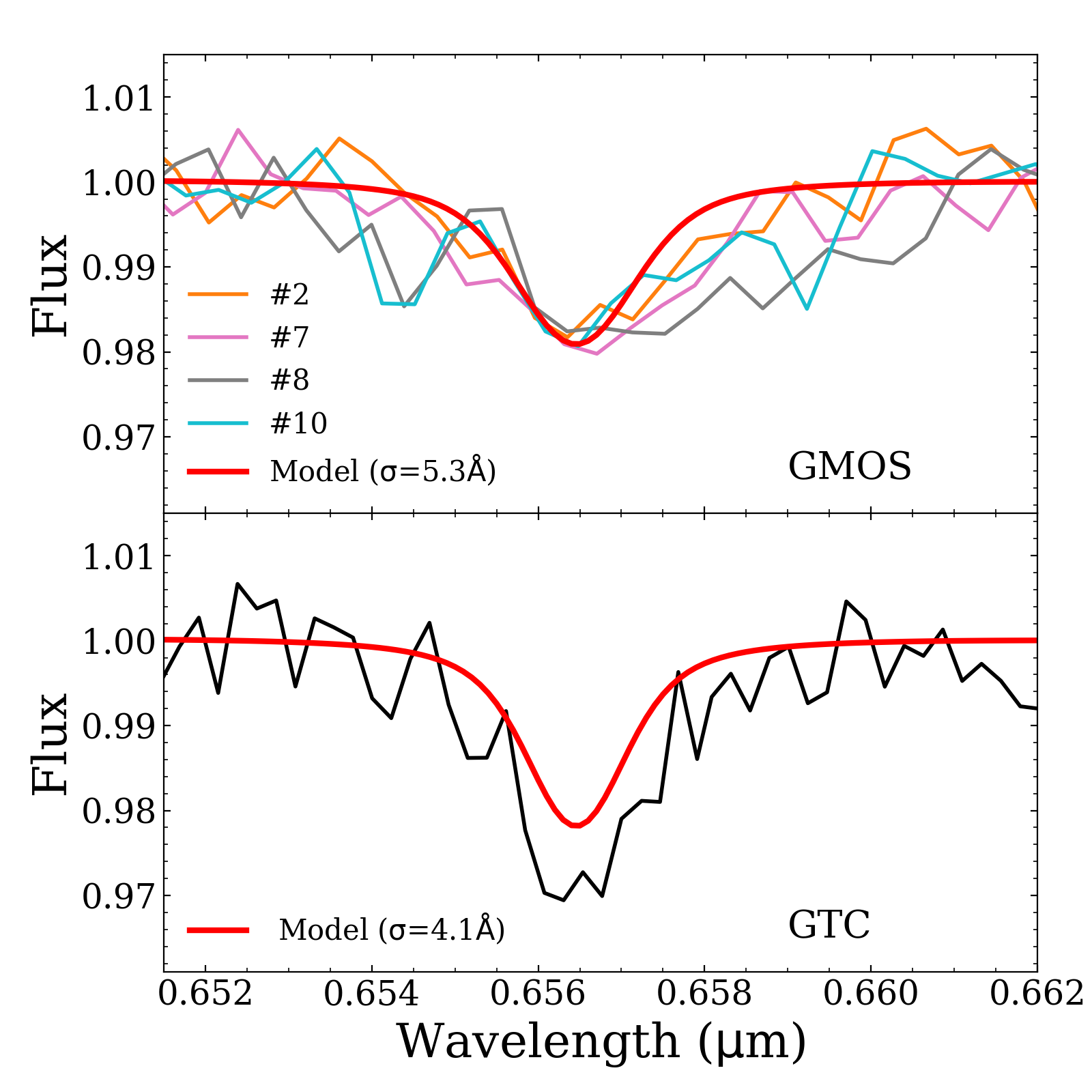}
\includegraphics[width=0.495\textwidth, trim={0.0cm 0.0cm 0.5cm 0.0cm}]{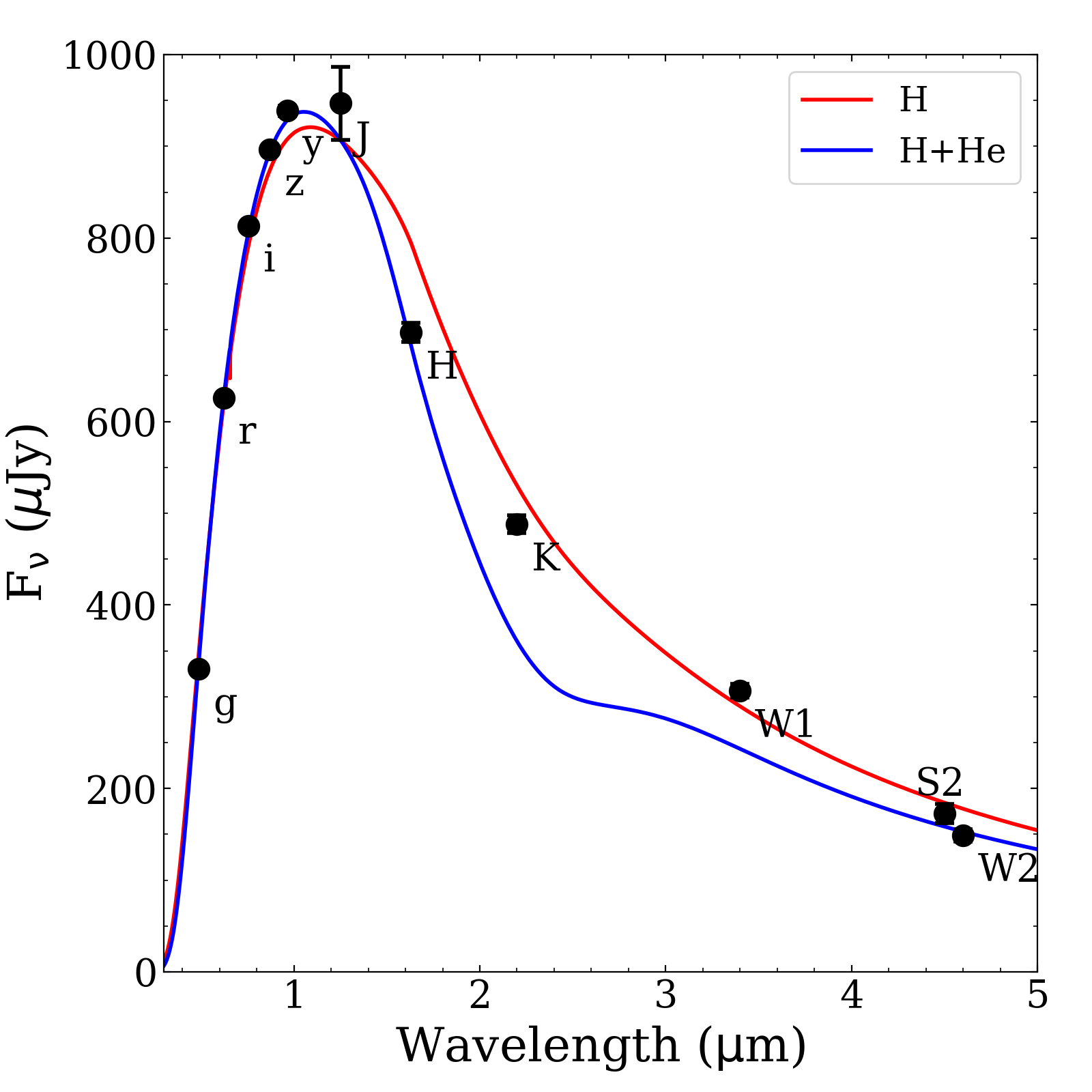}
\caption{{\it Left panel}: representative out-of-transit GMOS spectra of \object{WD~1856+534} and the GTC spectrum from \citet{Alonso2021}. The red solid line is the model convolved with a Gaussian profile to match the spectral resolution of the data. {\it Right panel}: SED fits to the photometry from Pan-Starrs, NIRI JHK, ALLWISE, and Spitzer 4.5~$\mu$m \citep{Vanderburg2020,Lai2021}. Red line is our best-fit H atmosphere model, while the blue line is the H+He mix atmosphere model in \citet{Vanderburg2020}.
}
\label{fig:SED}
\end{figure*}

We revisit the stellar parameters for \object{WD~1856+534} because they are essential to our analysis. \citet{Vanderburg2020} adopted a mixed atmosphere with 50\% hydrogen and 50\% helium due to the lack of H$\alpha$ in the optical spectrum from the Hobby-Eberly Telescope (HET). However, recent observations using the GTC detected a 3\% deep H$\alpha$ \citep{Alonso2021}. When we combine the out-of-transit spectra of \object{WD~1856+534} in each epoch, we have a tentative detection of H$\alpha$, as shown in Figure~\ref{fig:SED}. The GMOS spectra have a lower spectral resolution, but the wavelength and the strength of the feature is consistent with H$\alpha$ in the GTC spectrum. While the discrepancy between the HET nondetection of H$\alpha$ and the GTC and Gemini detections remains unexplained, one possibility is that the original HET spectrum may have had a calibration issue that over corrected the continuum region around H$\alpha$ (Zeimann, G. R., private communication). Our Gemini data confirm that \object{WD~1856+534} indeed is a DA white dwarf. 

To determine the white dwarf parameters, we performed a photometric fit following methods described in \citet{Vanderburg2020} with an updated JHK photometry of \object{WD~1856+534} \citep{Lai2021}. The white dwarf atmosphere was computed using the latest grids of cool white dwarf models \citep{Blouin2019}. The uncertainties were estimated by taking the difference between the best-fit solution assuming a minimum 0.03~mag uncertainty on the photometry (a common assumption, \citealt{Bergeron2019}) and the solution found using the nominal uncertainties. The best fit parameters are listed in Table~\ref{tab:properties} and the spectral energy distribution (SED) is shown in Figure~\ref{fig:SED}. An H atmosphere model simultaneously provides a good match to the SED and the H-alpha feature in the GTC and GMOS spectra. We found that this cannot be achieved with a H+He mix atmosphere model. In addition, we revised the upper limits in the atmospheric calcium and magnesium abundances using the HET spectrum. We also calculated the intensity profile across the white dwarf surface following \citet{Gianninas2013}. Assuming a quadratic limb darkening law, we fit the intensity profile and derive the limb darkening coefficients, as listed in Table~\ref{tab:properties}.

\begin{deluxetable*}{lcc}
\tablecaption{ \label{tab:properties} Properties of \object{WD~1856+534}}
\tablehead{
\colhead{Parameter}& \colhead{Value}
}
\startdata
Effective Temperature, T$_\mathrm{eff}$ & 4860 $\pm$ 80 K\\
Surface Gravity, log g & 7.995 $\pm$ 0.065 cm s$^{-2}$\\
Mass, M$_\mathrm{s}$	& 0.576 $\pm$ 0.040 M$_\mathrm{\odot}$ \\
Radius, R$_\mathrm{s}$ & 0.01263 $\pm$ 0.00050 R$_\mathrm{\odot}$  \\
Cooling Age, t$_\mathrm{cool}$ & 6.60 $\pm$ 0.48 Gyr \\
Calcium Abundance, log n(Ca)/n(H) & $<$ -11.2 \\
Magnesium Abundance, log n(Mg)/n(H) & $<$ -8.1 \\
Limb darkening parameters\tablenotemark{$\dag$}, u$_\mathrm{1}$ & 0.019 $\pm$ 0.017\\
u$_\mathrm{2}$ & 0.394 $\pm$ 0.022\\
\enddata
\tablenotetext{$\dag$}{This is derived from white dwarf models for  0.523--0.883~$\mathrm{\mu}$m.}
\end{deluxetable*}

\section{Transmission Spectrum of WD~1856+534~b \label{sec:LCs}}
\subsection{Generating Light Curves \& Fitting \label{sec:LC_fitting}}

We generate light curves from the time-series spectra following steps outlined in \citet{Diamond-Lowe2020}. Given the increased scatter at longer wavelengths (further discussed at the end of this subsection), we focus on flux from 0.523--0.883~$\mathrm{\mu}$m for the white light curves. We also divide the time-series of the white dwarf and the reference star spectra into 22 spectroscopic bins, with a width of 0.2~$\mathrm{\mu}$m each, to yield 22 spectroscopic light curves. To model the light curve systematics and transit parameters, we employ the open-source packages \textit{exoplanet} \citep{exoplanet} and \textit{celerite2} \citep{celerite1,celerite2}. In all our fits, the systematics is modeled with a Gaussian Process (GP) using a Mat\'ern 3/2 kernel, which assumes that the neighboring points are more likely to be correlated than those farther away.

\begin{figure*}
\gridline{\fig{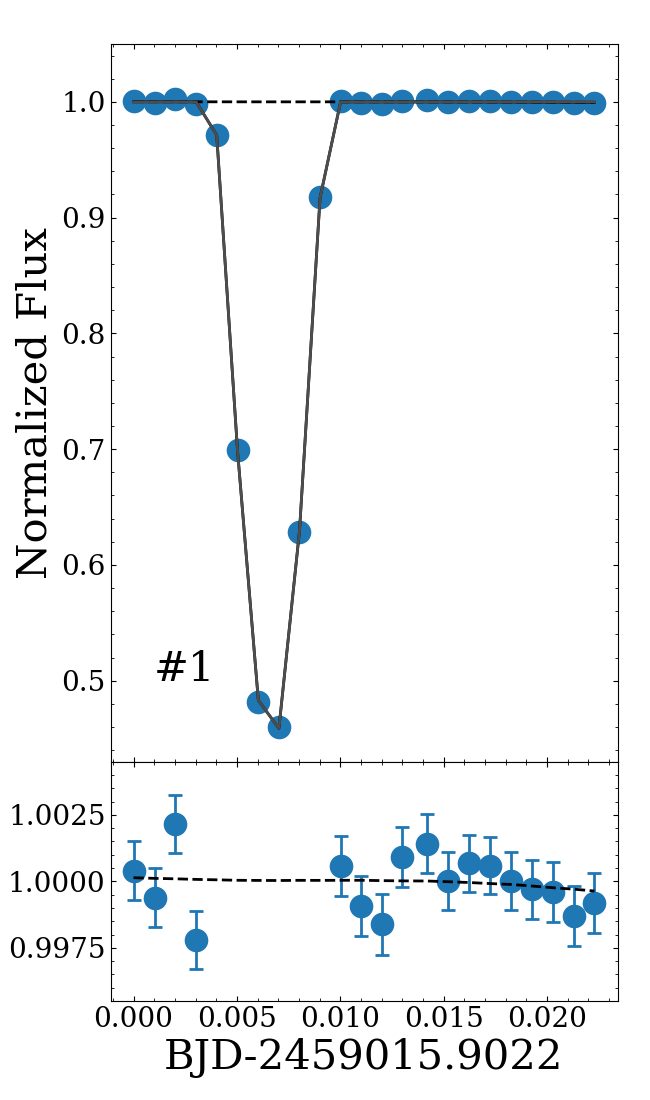}{0.2\textwidth}{}
\fig{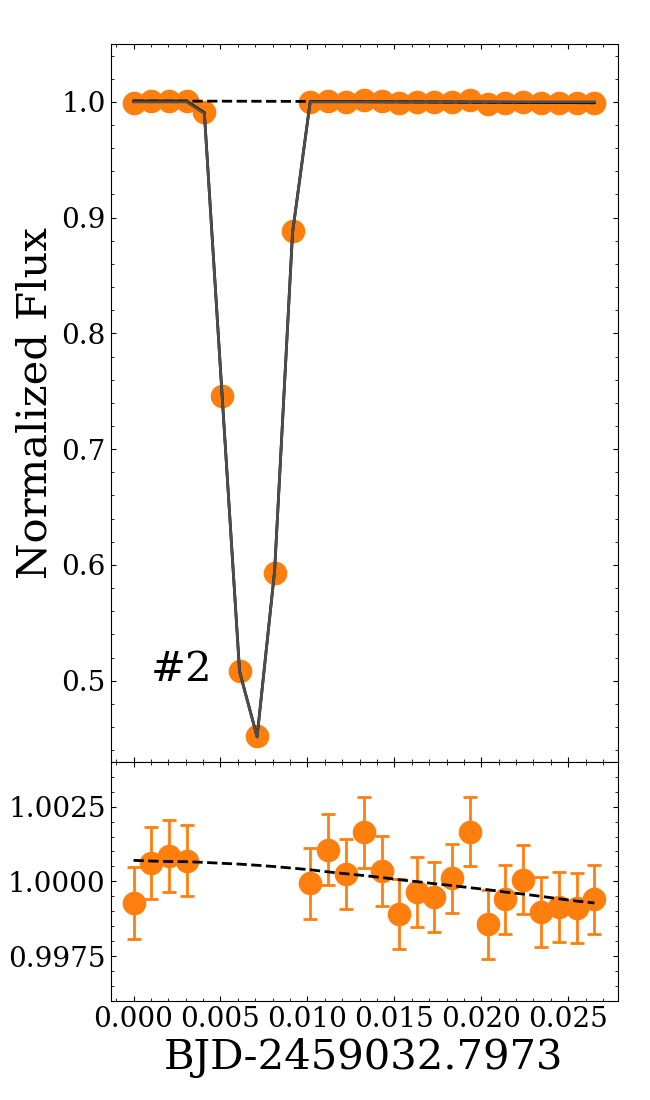}{0.2\textwidth}{}
\fig{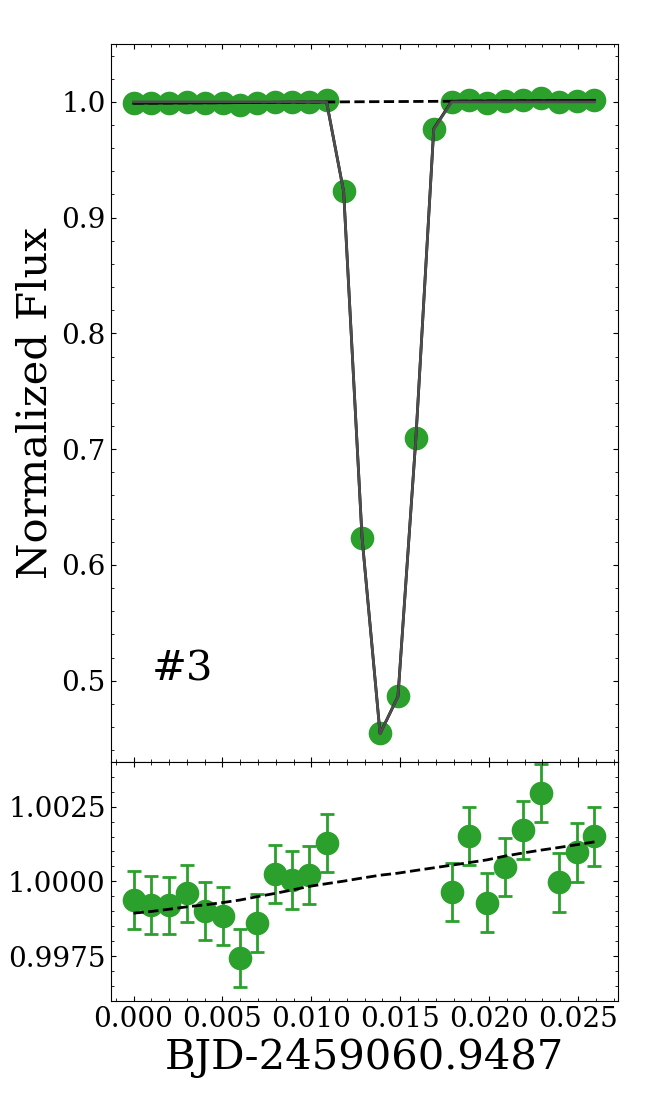}{0.2\textwidth}{}
\fig{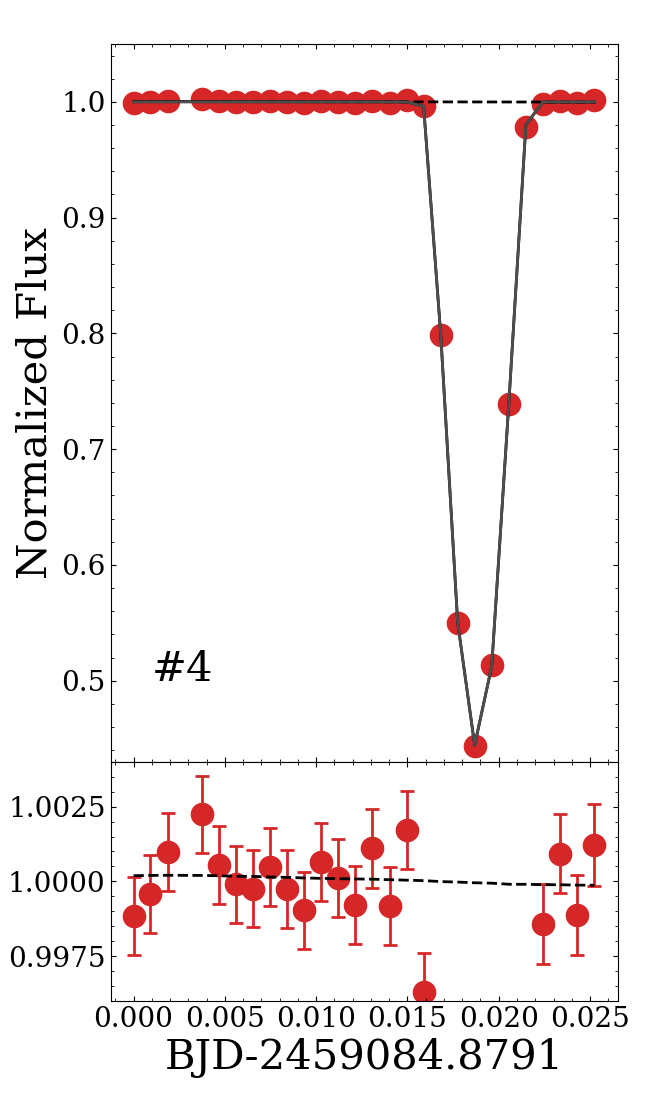}{0.2\textwidth}{}
\fig{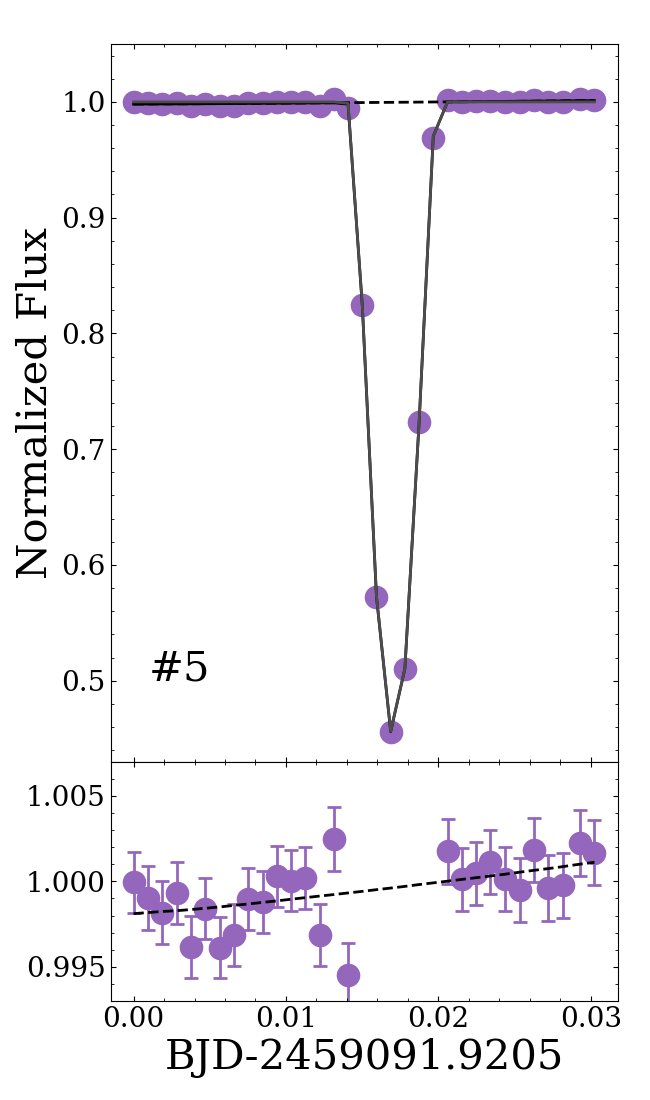}{0.2\textwidth}{}
}
\gridline{\fig{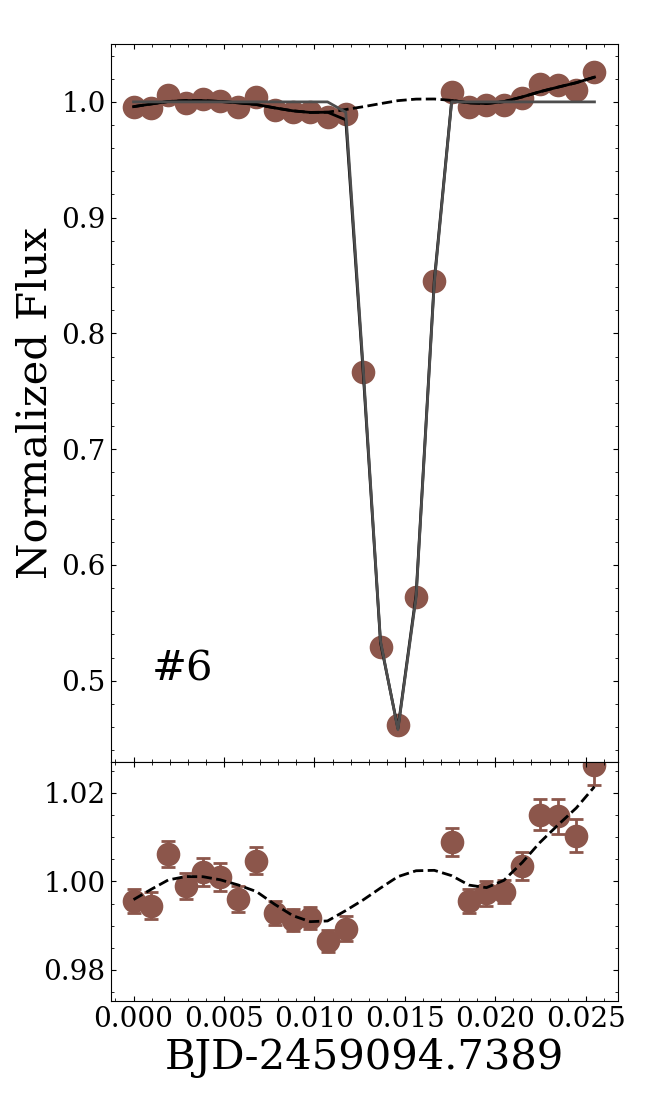}{0.2\textwidth}{}
\fig{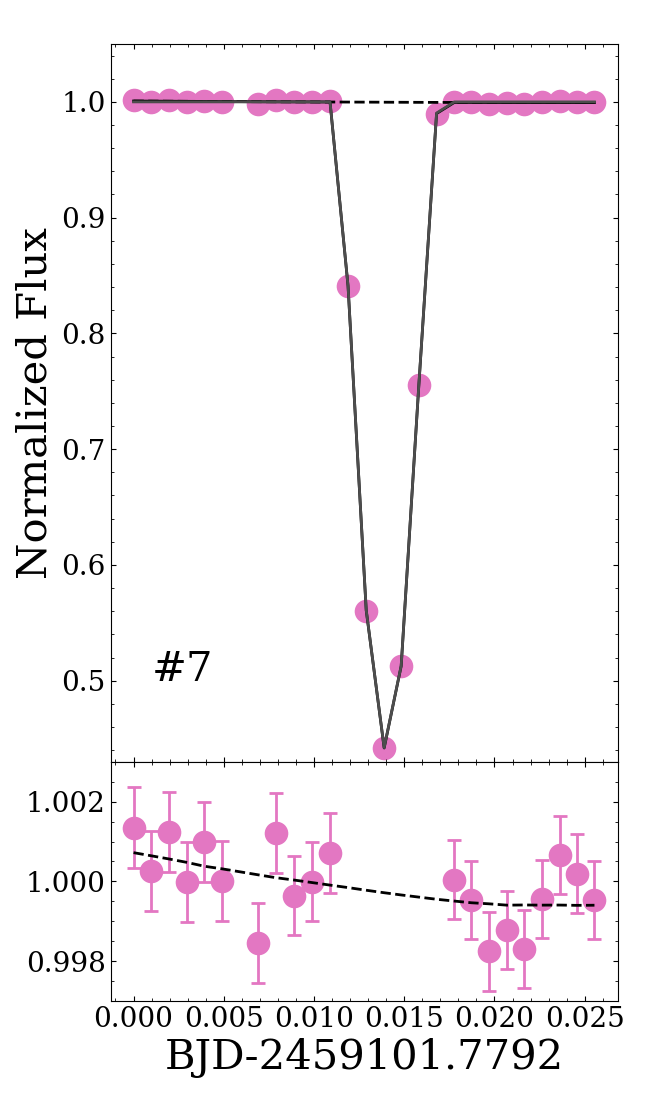}{0.2\textwidth}{}
\fig{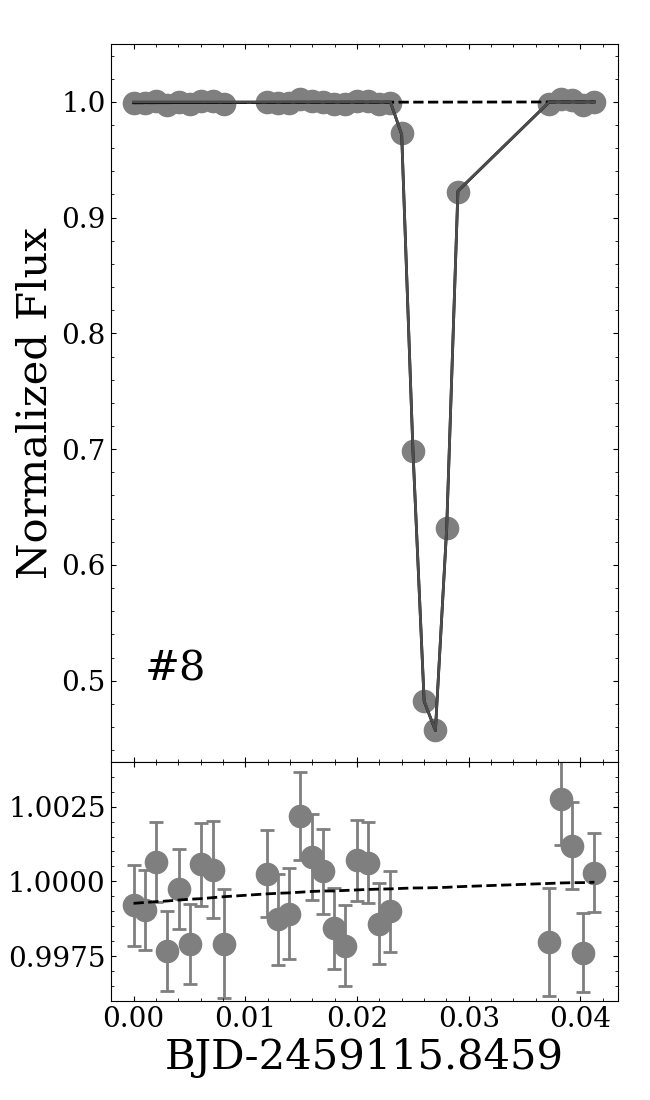}{0.2\textwidth}{}
\fig{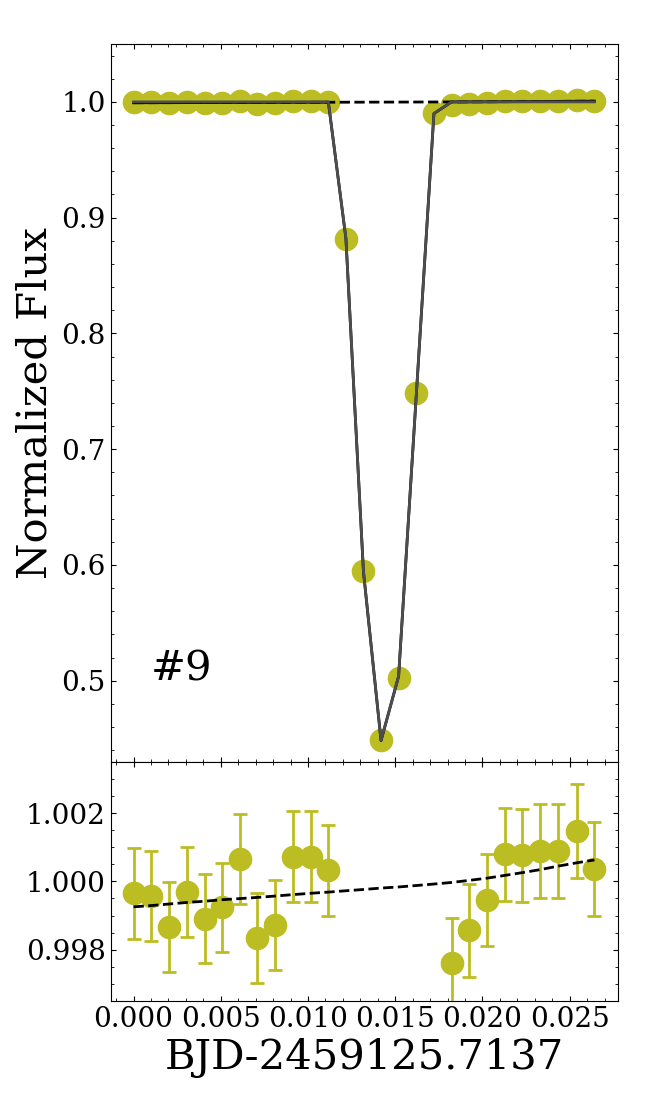}{0.2\textwidth}{}
\fig{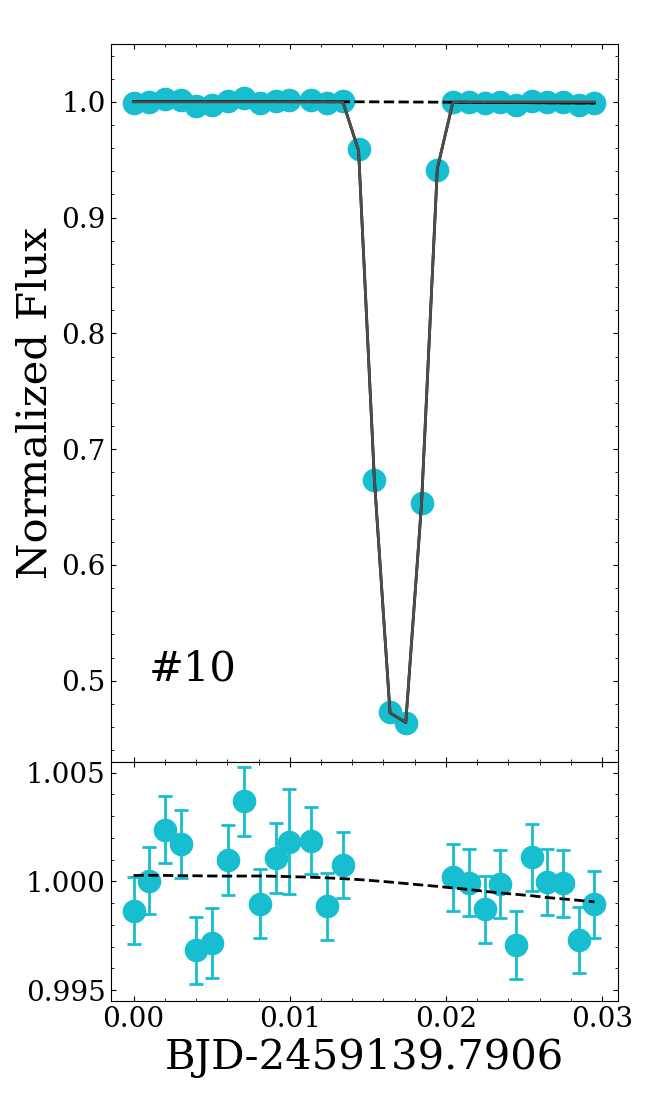}{0.2\textwidth}{}
}
\caption{Individual white light curves for our ten Gemini/GMOS transits of WD~1856+534~b. The bottom part of the figure is zoomed in around the out-of-transit light curve. At each epoch, there are only about six spectra taken during the 8~min transit. The solid lines represent the full result of the transit fit using \emph{exoplanet} and the dashed represents the GP fit to the systematics.
}
\label{fig:wLC}
\end{figure*}

We first identified the time of mid-transit, $t_0$, for each of our ten transit observations of WD~1856+534~b. We fit each white light curve separately, allowing $t_0$, the ratio of planet and stellar radii R$_\mathrm{p}$/R$_\mathrm{s}$, and the impact parameter $b$ to vary. We fix the values of stellar mass, stellar radius, and the limb darkening parameters to the values in Table~\ref{tab:properties}. The white light curves fitted with the best-fit transit and systematics models are shown in Figure~\ref{fig:wLC}. Transit \# 6 is especially noisy, likely due to the variable cloud conditions (CC80 as listed in Table~\ref{tab:log}). We derived the ephemeris in Table~\ref{tab:whiteLC} and it is consistent with those reported in \citet{Mallonn2020} and \citet{Alonso2021}.

To improve the time coverage, we phase-fold the white light curves by subtracting off the fitted mid-transit time for each individual epoch such that all transits are centered at time 0. Transits \#5 and \#6 were excluded from the analysis (see discussions below about their poor data quality). We now fit for R$_\mathrm{p}$/R$_\mathrm{s}$ and $b$ while simultaneously modeling any systematics. For the prior, we assume a uniform distribution for R$_\mathrm{p}$/R$_\mathrm{s}$ in 0.8--15.8 and $b$ in 1--10. All other transit parameters are fixed to the values in Table~\ref{tab:properties}. Compared to typical exoplanet transit observations, our observing duration is short and the data exhibit low levels of variability relative to the transit signal. We therefore choose to use a single systematic model for the phased white light curve. We sample the posteriors using PyMC3 Extras\footnote{\href{https://github.com/exoplanet-dev/pymc3-ext}{github.com/exoplanet-dev/pymc3-ext}} with 6 chains and 2000 draws. After sampling the posteriors we check that the chains are well mixed and have reached convergence. Our best-fit values for R$_\mathrm{p}$/R$_\mathrm{s}$ and $b$ from the phase-folded white light curve are listed in Table~\ref{tab:whiteLC}. We have reached a precision of 0.12~\% in the phase folded white light curve.

\begin{figure*}
\centering
\includegraphics[width=0.42\textwidth, trim={0.0cm 0.0cm 0.0cm 0.2cm}]{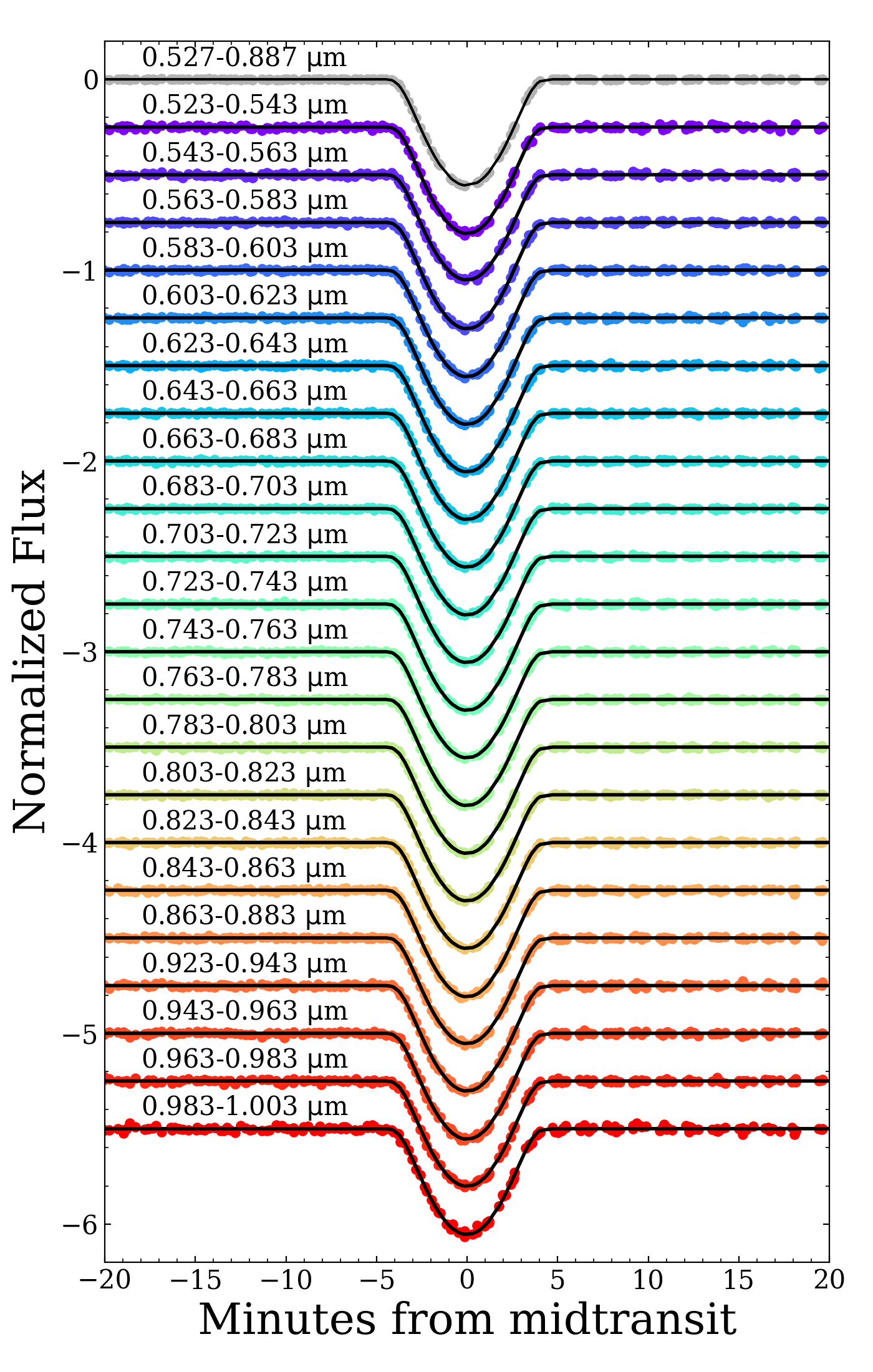}
\includegraphics[width=0.42\textwidth, trim={0.0cm 0.0cm 0.0cm 0.0cm}]{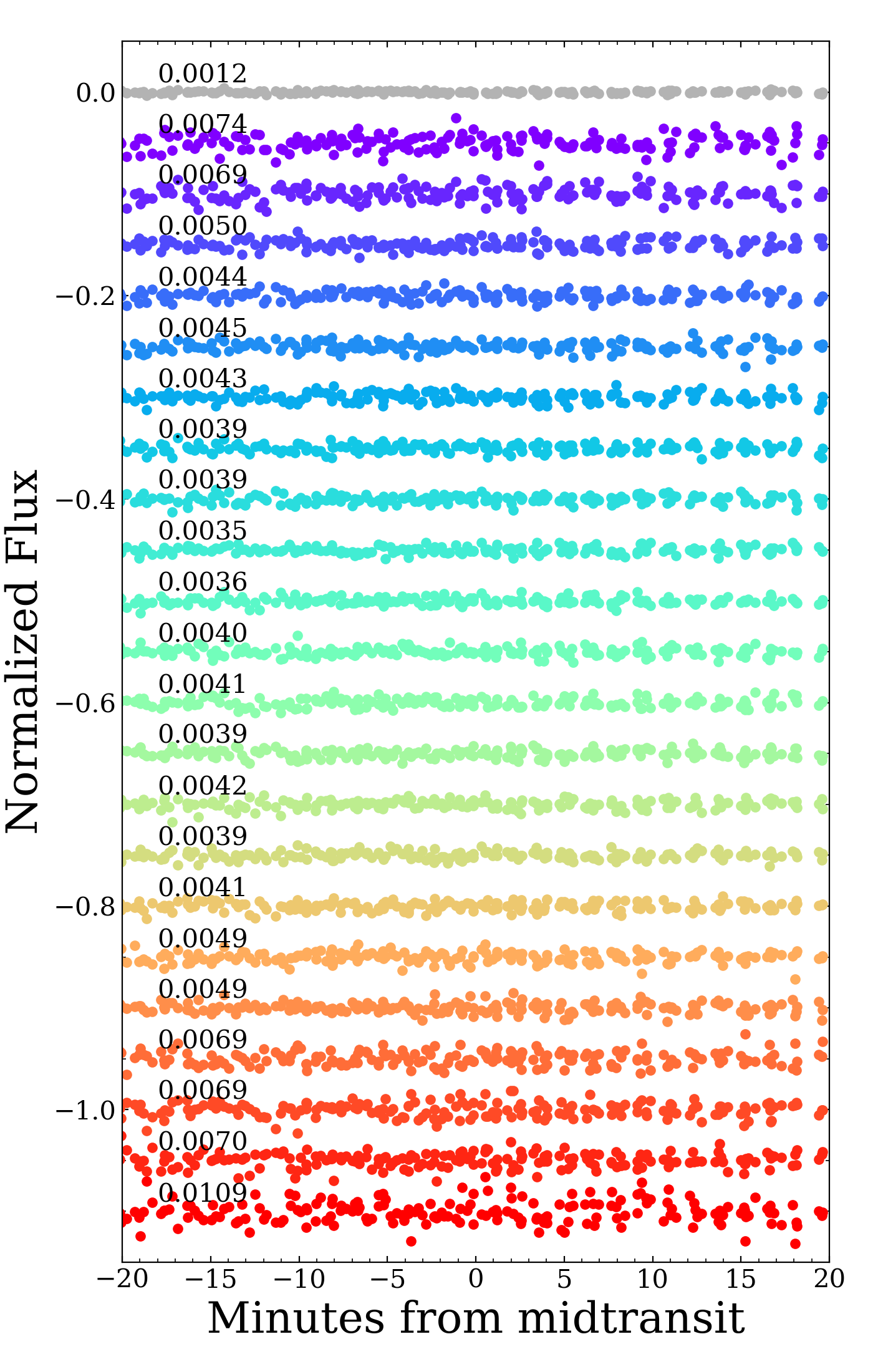}
\caption{Left panel: phase-folded light curves with the systematics removed in different wavelength bins (colored dots) and the best fit model (black lines). Right panel: residual from the fit. The standard deviations of the residuals are also annotated.
}
\label{fig:LCs}
\end{figure*}

Following a similar procedure for the white light curves, we also phase-fold the spectroscopic light curves. To fit the spectroscopic light curves, we include three transit parameters---the ratio of the planet and stellar radii R$_\mathrm{p}$/R$_\mathrm{s}$ and the resampled quadratic limb darkening parameters (q$_\mathrm{1}$, q$_\mathrm{2}$, \citealt{exoplanet:kipping13})---as well as the hyperparameters for the GP kernel. The impact parameter $b$ is fixed to the best-fit value from the phased white light curve. We sample the posteriors for each spectroscopic light curve using PyMC3 Extras as described for the phase-folded white light curve fit. The phase-folded spectroscopic data is shown in Figure~\ref{fig:LCs} and the best-fit parameters are listed in Table~\ref{tab:sLC}. The rms of the spectroscopic light curves is between 0.35\%--0.74\% with the best precision in the 0.643--0.863~$\mathrm{\mu}$m wavelength range.

\begin{figure*}
\centering
\includegraphics[width=\textwidth, trim={0.0cm 0.0cm 0.0cm 0.0cm}]{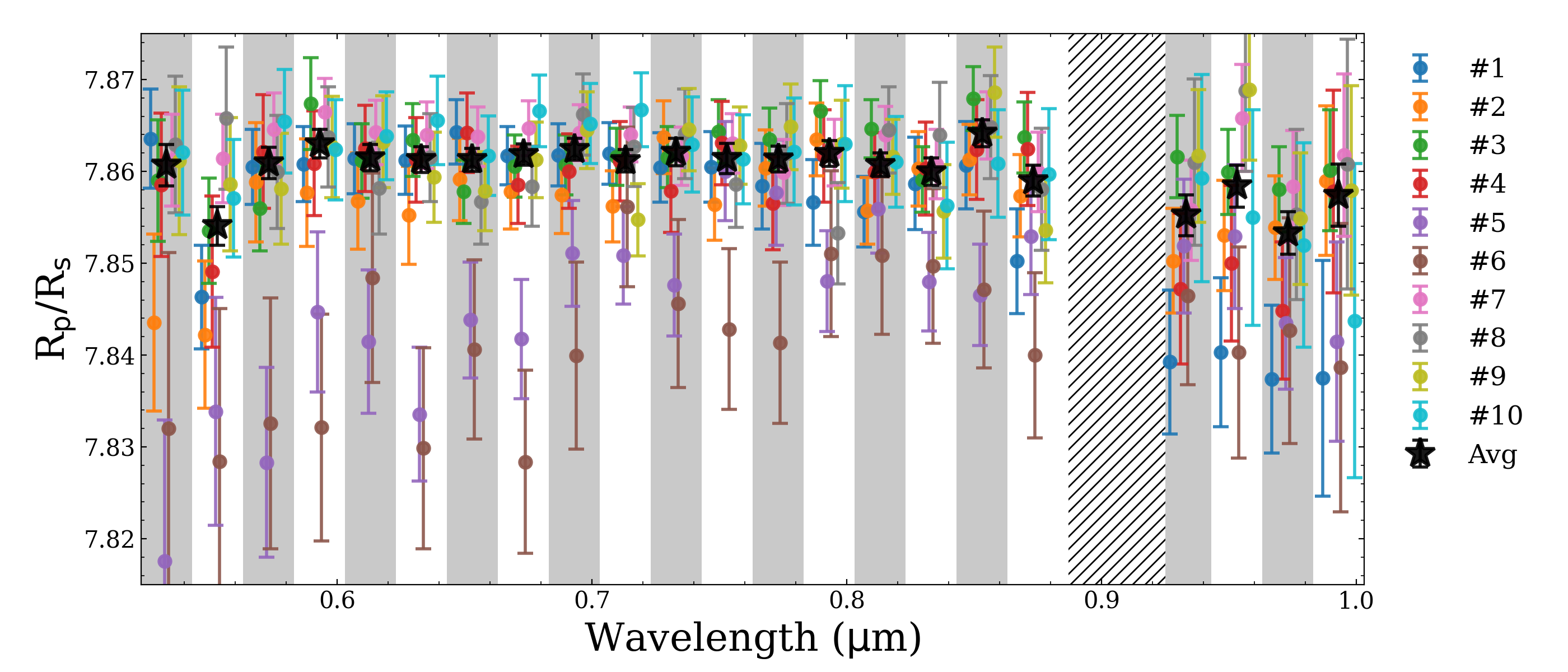}
\caption{The ratio of planet and stellar radii as a function of wavelength measured in each epoch (colored dots) and the phase-folded light curve (black stars, excluding transits \#5 and \#6). The grey and white bars are the spectral regions and the hatched area marks the detector chip gap. The results are mostly consistent among different epochs, except transits \#5 and \#6, which are likely affected by the variable sky background and therefore are excluded from the analysis.
}
\label{fig:rp/rwd}
\end{figure*}

To check the data quality in each individual epoch, we also fit the non-phase-folded spectroscopic light curves separately. The only free orbital parameter in this case is R$_\mathrm{p}$/R$_\mathrm{s}$. The limb darkening parameters are fixed to the values listed in Table~\ref{tab:sLC}. As shown in Figure~\ref{fig:rp/rwd}, the results are generally consistent among different epochs, demonstrating the long-term stability of GMOS. However, the values for R$_\mathrm{p}$/R$_\mathrm{s}$ obtained using transits \#5, \#6, and \#2 to a lesser extent, are systematically lower than the other seven transits. We speculate that this is caused by imperfect sky subtraction, particularly because these epochs have high sky background (all have sky background 100\%, as listed in Table~\ref{tab:log}). In addition, transit \#6 has the lowest overall counts due to the high cloud coverage. Therefore, we excluded transits \#5 and \#6 from the analysis. Figure~\ref{fig:rp/rwd} also shows that the overall scatter is much larger for wavelength bins beyond 0.9~$\mathrm{\mu}m$, possibly due to a combination of fringing and variable sky background at longer wavelengths \citep{Huitson2017}. For the rest of the paper, we only use 0.523--0.883~$\mathrm{\mu}$m wavelength range.

\begin{deluxetable}{lc}
\tablecaption{ \label{tab:whiteLC} White light curve properties}
\tablehead{
\colhead{Parameter} & \colhead{Value}
}
\startdata
Time of mid-transit, $t_0$ & 2459015.9088691 $\pm$ 0.0000022
BJD$_\mathrm{TDB}$ \\
Orbital period, P & 1.40793925 $\pm$ 0.00000004 days \\
Impact parameter, $b$ & 7.75 $\pm$ 0.01 \\
Radius ratio, R$_\mathrm{p}$/R$_\mathrm{s}$ & 7.86 $\pm$ 0.01 \\
\enddata
\end{deluxetable}

\begin{deluxetable}{clcccccccccc}
\tablecaption{ \label{tab:sLC} Derived transmission spectrum of WD~1856+534 b}
\tablehead{
\colhead{$\lambda$ ($\mathrm{\mu}$m)} & \colhead{q$_\mathrm{1}$}  & \colhead{q$_\mathrm{2}$} & \colhead{R$_\mathrm{p}$/R$_\mathrm{s}$} & \colhead{$\bar{\delta}_{\mathrm{corr}}$\tablenotemark{$\dagger$}}
}
\startdata
0.523--0.543 & 0.22$\pm$0.15 & 0.18$\pm$0.23 & 7.8607$\pm$0.0022 & 0.3284$\pm$0.0012 \\ 
0.543--0.563 & 0.09$\pm$0.07 & 0.06$\pm$0.10 & 7.8541$\pm$0.0021 & 0.3248$\pm$0.0011 \\ 
0.563--0.583 & 0.20$\pm$0.13 & 0.14$\pm$0.19 & 7.8609$\pm$0.0017 & 0.3284$\pm$0.0010 \\ 
0.583--0.603 & 0.24$\pm$0.13 & 0.07$\pm$0.19 & 7.8630$\pm$0.0016 & 0.3296$\pm$0.0009 \\ 
0.603--0.623 & 0.24$\pm$0.14 & 0.10$\pm$0.20 & 7.8615$\pm$0.0014 & 0.3287$\pm$0.0007 \\ 
0.623--0.643 & 0.21$\pm$0.12 & 0.07$\pm$0.16 & 7.8613$\pm$0.0014 & 0.3287$\pm$0.0007 \\ 
0.643--0.663 & 0.14$\pm$0.10 & 0.11$\pm$0.14 & 7.8613$\pm$0.0012 & 0.3286$\pm$0.0007 \\ 
0.663--0.683 & 0.10$\pm$0.07 & 0.08$\pm$0.11 & 7.8618$\pm$0.0011 & 0.3290$\pm$0.0006 \\ 
0.683--0.703 & 0.15$\pm$0.09 & 0.06$\pm$0.13 & 7.8624$\pm$0.0012 & 0.3294$\pm$0.0007 \\ 
0.703--0.723 & 0.13$\pm$0.09 & 0.10$\pm$0.13 & 7.8612$\pm$0.0012 & 0.3286$\pm$0.0006 \\ 
0.723--0.743 & 0.15$\pm$0.10 & 0.08$\pm$0.14 & 7.8621$\pm$0.0015 & 0.3291$\pm$0.0008 \\ 
0.743--0.763 & 0.09$\pm$0.07 & 0.06$\pm$0.09 & 7.8614$\pm$0.0016 & 0.3288$\pm$0.0009 \\ 
0.763--0.783 & 0.13$\pm$0.09 & 0.09$\pm$0.13 & 7.8613$\pm$0.0014 & 0.3287$\pm$0.0008 \\ 
0.783--0.803 & 0.13$\pm$0.09 & 0.08$\pm$0.13 & 7.8620$\pm$0.0014 & 0.3290$\pm$0.0007 \\ 
0.803--0.823 & 0.11$\pm$0.07 & 0.10$\pm$0.11 & 7.8608$\pm$0.0013 & 0.3284$\pm$0.0007 \\ 
0.823--0.843 & 0.11$\pm$0.08 & 0.05$\pm$0.10 & 7.8600$\pm$0.0015 & 0.3279$\pm$0.0008 \\ 
0.843--0.863 & 0.16$\pm$0.09 & 0.04$\pm$0.12 & 7.8641$\pm$0.0015 & 0.3302$\pm$0.0008 \\ 
0.863--0.883 & 0.09$\pm$0.07 & 0.04$\pm$0.08 & 7.8590$\pm$0.0017 & 0.3273$\pm$0.0009 \\ 
0.923--0.943\tablenotemark{$\ddag$} & 0.14$\pm$0.10 & 0.08$\pm$0.14 & 7.8552$\pm$0.0022 & 0.3254$\pm$0.0012 \\ 
0.943--0.963\tablenotemark{$\ddag$} & 0.14$\pm$0.10 & 0.08$\pm$0.13 & 7.8584$\pm$0.0023 & 0.3271$\pm$0.0013 \\ 
0.963--0.983\tablenotemark{$\ddag$} & 0.14$\pm$0.10 & 0.08$\pm$0.14 & 7.8533$\pm$0.0023 & 0.3243$\pm$0.0013 \\ 
0.983--1.003\tablenotemark{$\ddag$} & 0.15$\pm$0.12 & 0.07$\pm$0.14 & 7.8574$\pm$0.0034 & 0.3266$\pm$0.0019 \\ 
  \enddata
\tablenotetext{$\dag$}{This is the limb darkening corrected time-averaged transit depth.}
\tablenotetext{$\ddag$}{These bins are excluded from the atmosphere fitting analysis, see discussion in Section~\ref{sec:LC_fitting}.}
\tablecomments{Table~\ref{tab:sLC} is published in the machine-readable format.}
\end{deluxetable}

\subsection{A limb darkening corrected, time-averaged transmission spectrum} \label{subsec:grazing_transit_spectra}

\begin{figure*}[ht!]
\centering
\includegraphics[width=\textwidth, trim={0.0cm 0.0cm 0.0cm 0.0cm}]{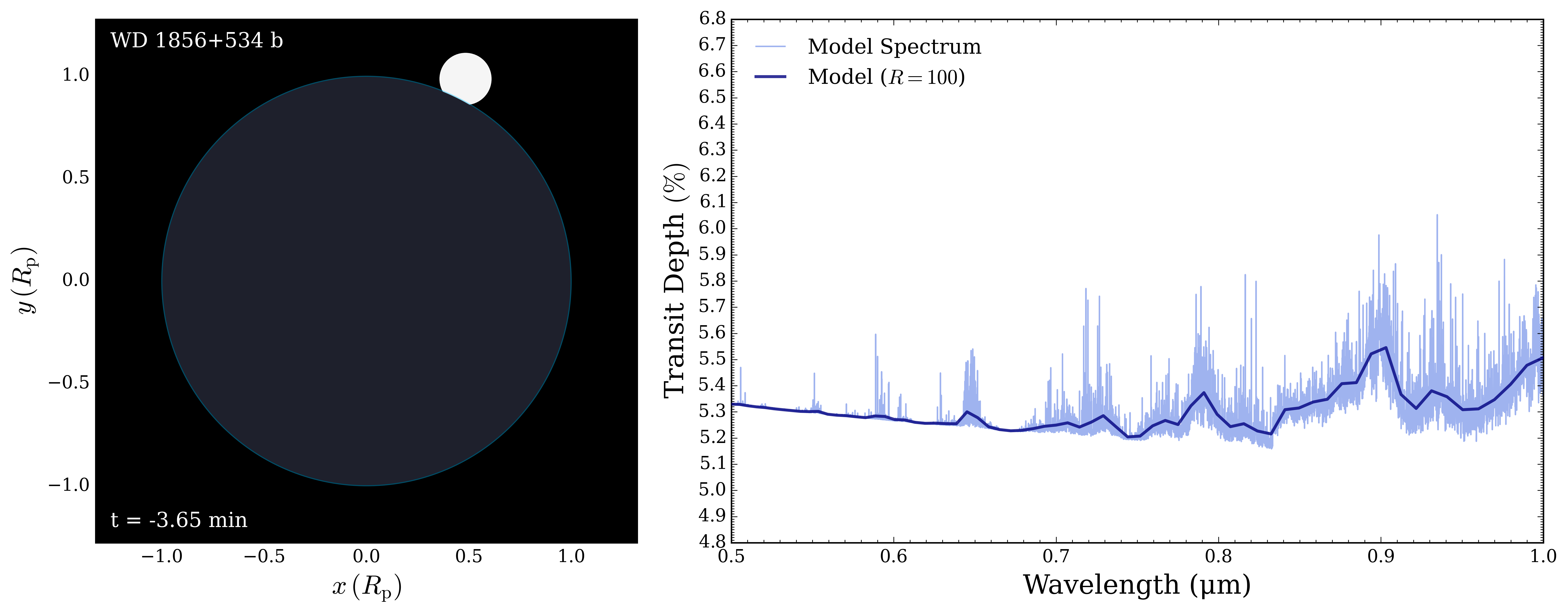}
\includegraphics[width=\textwidth, trim={0.0cm 0.0cm 0.0cm 0.0cm}]{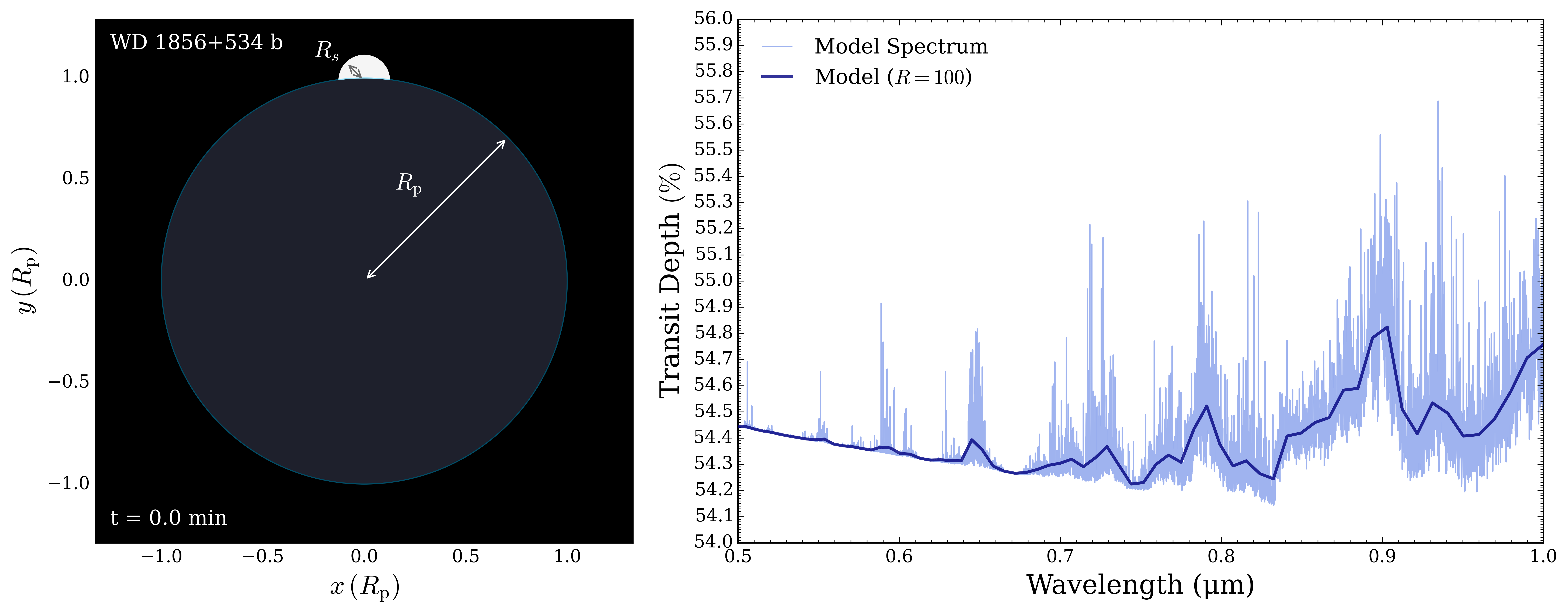}
\includegraphics[width=\textwidth, trim={0.0cm 0.0cm 0.0cm 0.0cm}]{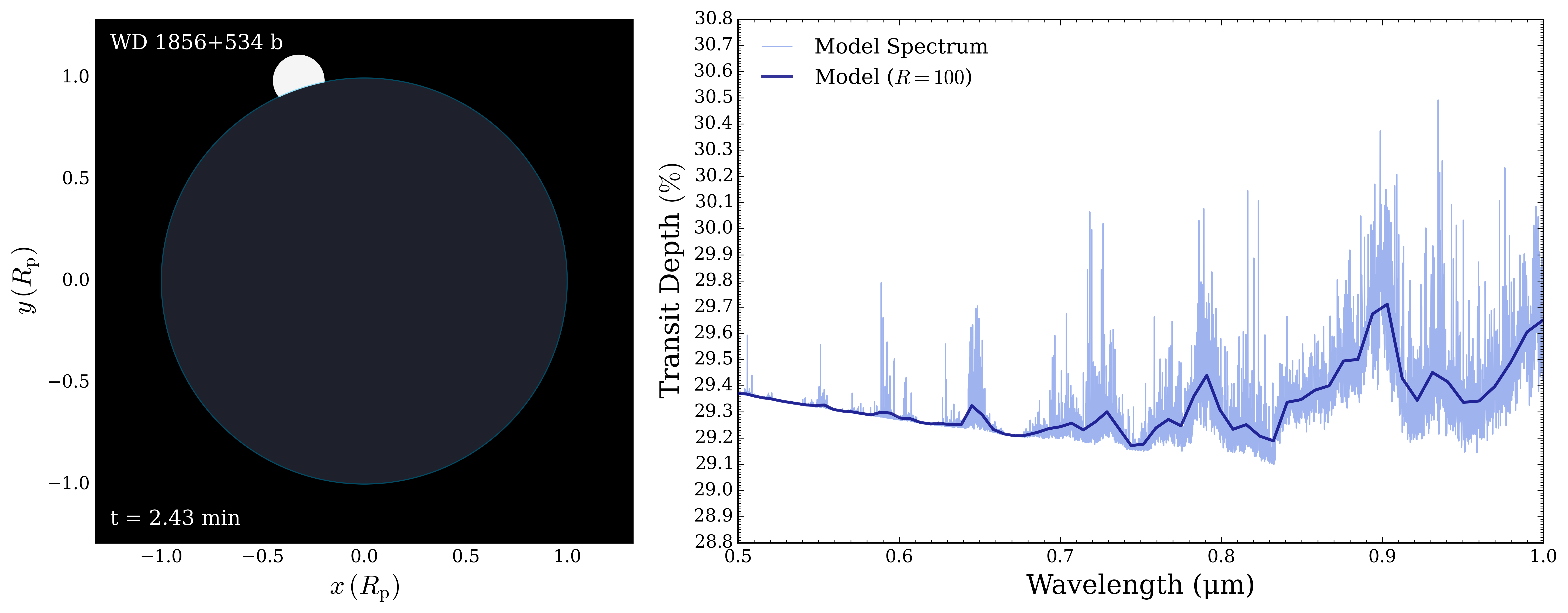}
\caption{Animation of a WD~1856+534~b transit and the associated transmission spectrum. An animated version of this figure is available. The animation begins at $t = -6.08$ minutes and ends at $t = 6.0$ minutes. The realtime duration of the animation is 10 seconds.
The static figure shows three frames sampling the ingress, mid-transit, and egress. Left: system geometry during a transit, showing the white dwarf (white circle), WD~1856+534~b (grey circle), and the planetary atmosphere (blue annulus) to scale. Right: a model transmission spectrum of WD~1856+534~b, assuming a mass of 1\,$M_J$ and a Jovian composition, at the same time steps as each left panel. The absorption feature amplitudes and Rayleigh slope in the transmission spectrum vary with time throughout the transit due to the changing fraction of the planetary atmosphere occulting the white dwarf.  }
\label{fig:transit_animation}
\end{figure*}

The transit geometry of the WD~1856+534~b system, with a planet larger than its star and a highly grazing transit, requires a careful reconsideration of standard approaches to transmission spectroscopy. For typical transits, the usual analysis procedure is as follows: (i) produce spectroscopic light curves and fit them with a systematics and transit model; (ii) extract the posterior distribution of the ratio of the planet to stellar radii $R_{\mathrm{p}} / R_{\mathrm{s}}$ at each wavelength; (iii) compare $(R_{\mathrm{p}, \lambda} / R_{\mathrm{s}})^2$ to model transmission spectra. However, the last step relies on a critical assumption: the transit depth of a planet occulting a star with a uniform intensity distribution is equal to $(R_{\mathrm{p, eff}, \lambda} / R_{\mathrm{s}})^2$ (i.e. the effective area ratio between the planet and star). This assumption does not hold for WD~1856+534~b---or for grazing transits more generally---since the fraction of the planetary disc, and its atmosphere, occulting the star changes throughout the transit. Furthermore, different regions of WD~1856+534~b's atmosphere are probed throughout a transit. We illustrate these critical points via an animation in Figure~\ref{fig:transit_animation}. So while one can extract a spectrum of $(R_{\mathrm{p}, \lambda} / R_{\mathrm{s}})^2$ from light curves, the equivalence between this quantity and model transmission spectra breaks down for grazing transits. An alternative quantity is needed to compare spectroscopic transit observations and model transmission spectra.

\begin{figure*}
\centering
\includegraphics[width=\textwidth]{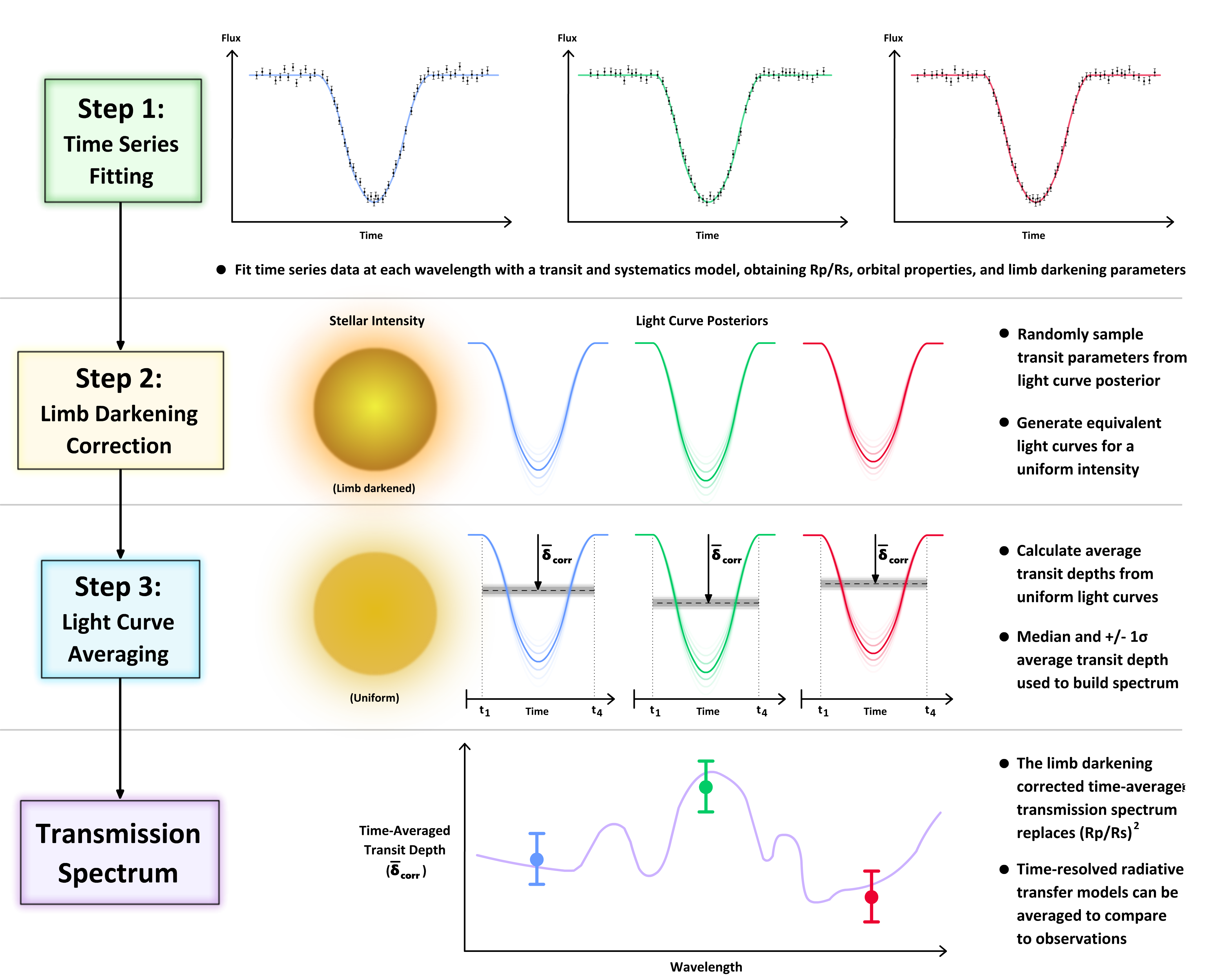}
\caption{Our proposed procedure to extract the transmission spectrum of a planet with a grazing transit. The steps are described in detail in Section~\ref{subsec:grazing_transit_spectra}.}
\label{fig:grazing_transit_spectrum_extraction}
\end{figure*}

Here, we introduce the concept of the \emph{limb darkening corrected, time-averaged transmission spectrum} as a suitable replacement for $(R_{\mathrm{p}, \lambda} / R_{\mathrm{s}})^2$ in the context of grazing transits. Figure~\ref{fig:grazing_transit_spectrum_extraction} illustrates the definition of this quantity and how it is extracted from time series transit observations. In step 1, identical to the standard approach, spectroscopic time series data is fit by a combined transit and systematics model. In step 2, the posterior distributions from the light curve fits are randomly sampled to extract the parameter distributions underlying the transit model ($R_{\mathrm{p}}/R_{\mathrm{s}}$, orbital properties, limb darkening parameters, etc). These parameters are then fed back through the transit forward model to compute the \emph{equivalent} light curves for a uniform stellar intensity distribution (here, $q_1$ and $q_2$ are set to zero at all wavelengths). These limb darkening corrected light curves represent how the planetary transit would appear if the star had a uniform intensity distribution. In step 3, an array of transit depths are computed as a function of time from each limb darkening corrected light curve. The mean value of this array between first and fourth contact is the `limb darkening corrected time-averaged transit depth', which we denote as ${\bar{\delta}_{\mathrm{corr}}}$. After computing $\bar{\delta}_{\mathrm{corr}}$ for all the random samples drawn from the original time series fit, we take the median time-averaged transit depth at each wavelength, $\bar{\delta}_{\mathrm{corr}} (\lambda)$, as our transmission spectrum. The associated 1\,$\sigma$ distribution for $\bar{\delta}_{\mathrm{corr}} (\lambda)$ represents the uncertainty in the observed transmission spectrum. Crucially, theoretical $\bar{\delta}_{\mathrm{corr}} (\lambda)$ can be readily computed from radiative transfer codes (see Section~\ref{sec:theory_models_WD1856b}) to compare with observed transmission spectra.

We applied this new method to our observations of WD~1856+534~b. The procedures for time series fitting (step 1) are described in Section~\ref{sec:LC_fitting}. For step 2, we generated 500 model light curves in each spectroscopic bin from the posterior distribution of the radius ratio, R$_\mathrm{p}$/R$_\mathrm{s}$, for a uniform stellar intensity distribution (i.e., q$_\mathrm{1}$=0, q$_\mathrm{2}$=0). The other transit parameters are taken from Table~\ref{tab:properties}. To calculate the limb darkening corrected time-averaged transit depth, $\bar{\delta}_\mathrm{corr}$ (step 3), we sampled the model light curves at 1000 evenly spaced points between orbital phases of -0.002 and 0.002 (T1 and T4, respectively). Our final derived transmission spectrum is presented in Table~\ref{tab:sLC}, both in terms of $R_{\mathrm{p}} / R_{\mathrm{s}}$ and $\bar{\delta}_{\mathrm{corr}}$.

\section{Radiative Transfer and Atmospheric Modeling for WD~1856+534~b} \label{sec:theory_models_WD1856b}

\subsection{Radiative Transfer} \label{subsec:radiative_transfer}

A model transmission spectrum can be expressed, assuming a uniform stellar intensity, as the ratio of the effective area of the occulting planet to that of its host star. In the usual case of a planet completely occulting the stellar disc, this leads to a standard expression for the wavelength-dependent transit depth
\begin{equation}
\delta_{\lambda} = \left(\frac{R_{\mathrm{p, \, eff, \, \lambda}}}{{R_{s}}}\right)^{2} = \frac{R_{\mathrm{p, \, top}}^{2} - \displaystyle\int_{0}^{R_{\mathrm{top}}} 2 \rho \, e^{-\tau_{\lambda}(\rho)} \, d\rho}{R_{s}^2}
\label{eq:transit_depth_complete_overlap}
\end{equation} 
where $R_{\mathrm{p, \, top}}$ is the top radial limit of the modeled atmosphere, $\tau_{\lambda}$ is the slant optical depth, and $\rho$ is the impact parameter of a given ray. The first term gives the transit depth if the entire atmosphere is opaque, while the second term subtracts the area of atmospheric annuli weighted by the transmission ($e^{-\tau_{\lambda}}$) of each layer. Given the assumption of a uniform stellar intensity, a translation of the planet to a different location on the stellar disc leaves the transmission spectrum unchanged. The model spectrum is, therefore, time-invariant during transit and one is free to pick the reference time when the spectrum is computed (normally mid-transit).

For a grazing transit, a more general expression must be used. A suitable generalization is given by
\begin{equation}
\delta_{\lambda} (t) = \frac{A_{\mathrm{p, \, overlap}} (t) -  \displaystyle\int_{A_{\mathrm{p, \, overlap}} (t)} e^{-\tau_{\lambda}} \, dA}{\pi R_{s}^2}
\label{eq:transit_depth_grazing}
\end{equation} 
where $A_{\mathrm{p, \, overlap}}$ is the combined area of the planetary disc and atmosphere occulting the star and $t$ denotes the time during transit. $A_{\mathrm{p, \, overlap}}$ is given by the area of overlap between two circles of radii $R_{\mathrm{p, \, top}}$ and $R_{s}$ at a distance specified by the impact parameter, $b$, and the time, $t$. The integral can be numerically computed over the overlapping area, with the transmission multiplied by area elements. The azimuthal integral no longer trivially evaluates to $2 \pi$ (even for a 1D atmosphere), as in Equation~\ref{eq:transit_depth_complete_overlap}, since only area elements where the atmosphere occults the stellar disc contribute to the integral. Consequently, the transmission spectrum itself is \emph{time-dependent} throughout the transit.

We illustrate the system geometry and time-dependence of WD~1856+534~b's transmission spectrum via the animation in Figure~\ref{fig:transit_animation}. The transmission spectrum displays two key time-dependent features: (i) the changing bulk transit depth that reflects the changing opaque area of the planet disc; and (ii) the changing amplitude of absorption features as the fraction of the atmosphere covering the stellar disc changes. The latter effect results in a steeper Rayleigh slope at mid-transit than during ingress or egress. Given that the atmospheric regions probed by stellar photons change throughout the transit, a model computed at a single time step will not be representative of the transit as a whole. We hence conclude that a time-average of a series of model spectra(i.e. the average across all the frames in Figure~\ref{fig:transit_animation}) is a better model, allowing direct comparison with the limb darkening corrected time-averaged transmission spectra. 

We calculated time-resolved transmission spectra with the POSEIDON radiative transfer code \citep{MacDonald2017}, updated here to account for grazing transit geometries. We discretize the atmosphere into 100 layers (spaced uniformly in log-pressure from $10^{-7}$ to 10\,bar) and 720 azimuthal sectors ($0.5^{\circ}$ spacing). At each time step, we compute a model transmission spectrum at $R$ = 10,000 by sampling high-resolution molecular cross sections (pre-computed at $\Delta \nu =$ 0.01\,cm$^{-1}$). In this work, we use the latest ExoMol \citep{Tennyson2016} line lists for NH$_3$ \citep{Coles2019}, CH$_4$ \citep{Yurchenko2017}, and H$_2$O \citep{Polyansky2018}. We evaluate transmission spectra at 51 time steps from first contact through fourth contact (phases from -0.002 to 0.002) using Equation~\ref{eq:transit_depth_grazing}. The specific time steps are calculated using the \textit{batman} package \citep{Kreidberg2015}, accounting for the changing transverse orbital velocity of WD~1856+534~b throughout the transit. Finally, the spectra are time-averaged using trapezium rule integration to yield a model spectrum that can be directly compared with $\bar{\delta}_{\mathrm{corr}} (\lambda)$. 

\subsection{Atmospheric Models for WD~1856+534~b}

WD~1856+534~b is analogous to Jupiter, representing a cool gas giant nearly unique amongst the known transiting exoplanet population. At an equilibrium temperature of 163$^{+14}_{-18}$\,K \citep{Vanderburg2020}, WD~1856+534~b is marginally warmer than Jupiter's effective temperature (125\,K, \citealt{Ingersoll1976}). At this temperature, the dominant C, N, and O carriers in chemical equilibrium are CH$_4$, NH$_3$, and H$_2$O \citep{Woitke2018}. Remote and in-situ exploration of Jupiter has confirmed this picture \citep[e.g.][]{Atreya2003,Li2020}. The cool brown dwarf WISE 0855 ($\approx$ $250$\,K) also resembles a slightly warmer variant of Jupiter, showing CH$_4$ and H$_2$O absorption features \citep{Skemer2016,Morley2018}. To first order, we therefore expect the spectrum of WD~1856+534~b to be shaped by CH$_4$, NH$_3$, and H$_2$O, which all have absorption bands in the wavelength range covered by our GMOS spectrum.

\begin{figure*}
\centering
\includegraphics[width=0.49\textwidth, trim={0.0cm 0.0cm 0.0cm 0.0cm}]{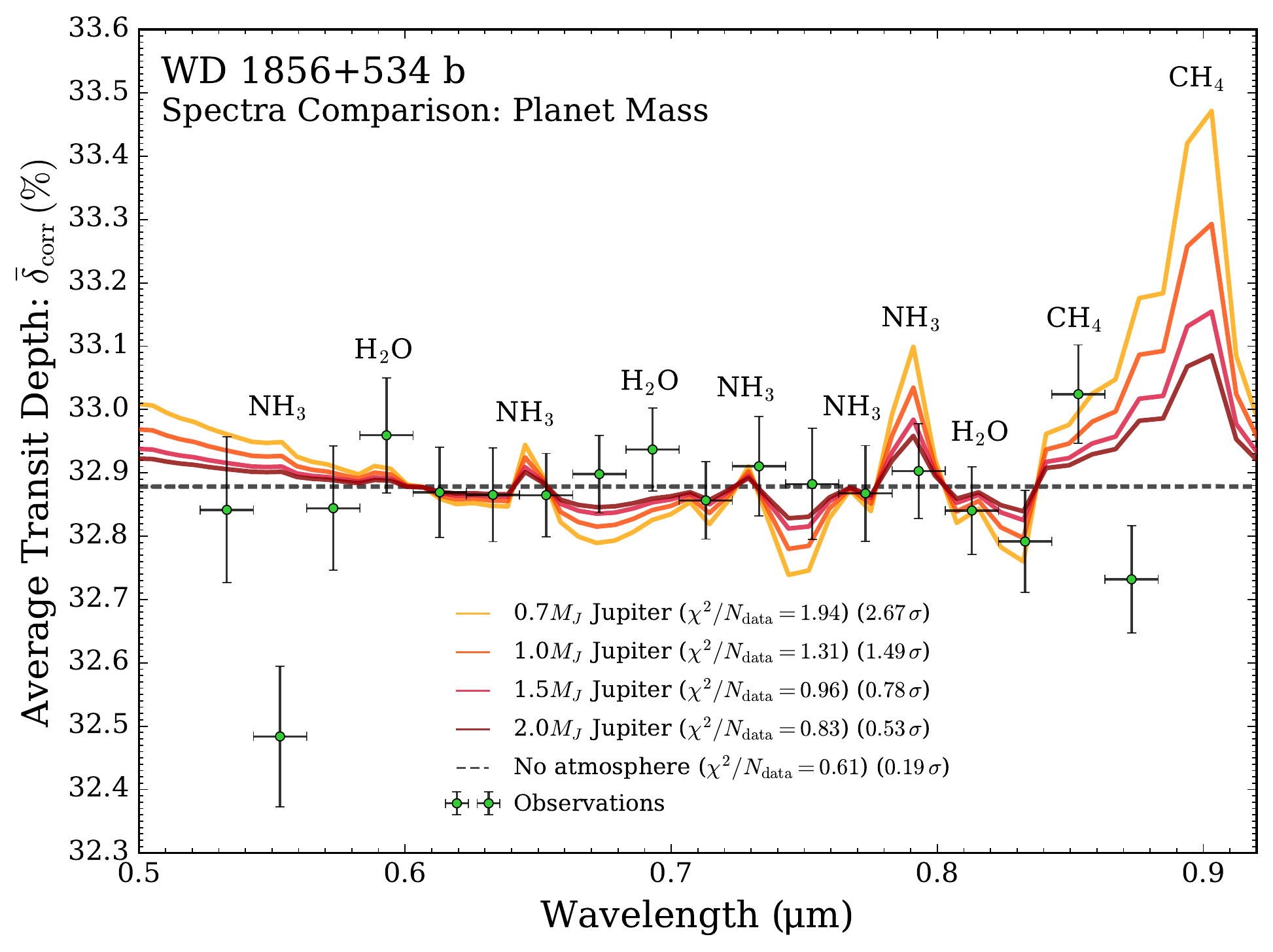}
\includegraphics[width=0.49\textwidth, trim={0.0cm 0.0cm 0.0cm 0.0cm}]{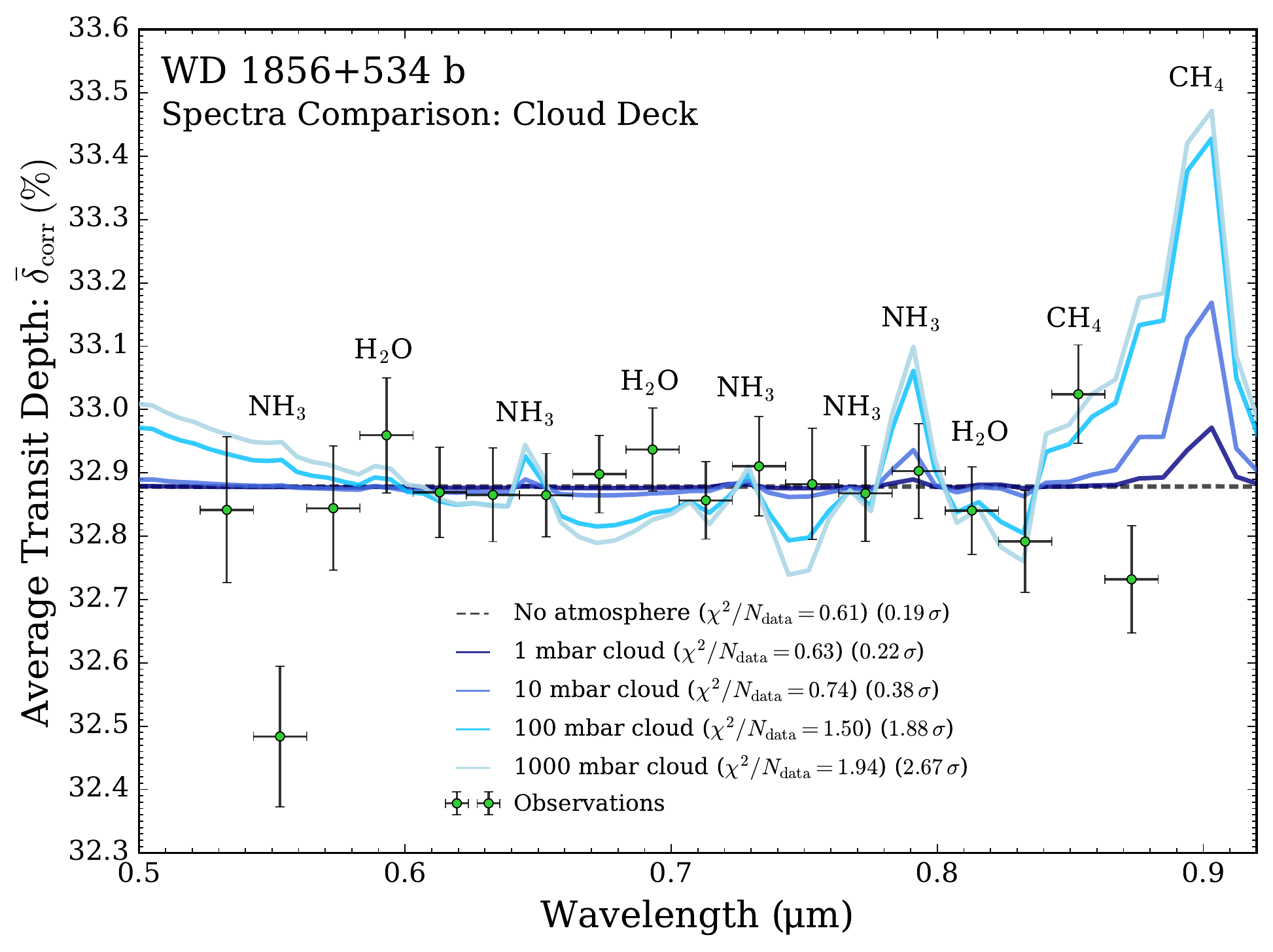}
\caption{Comparison of our Gemini/GMOS transmission spectrum of WD~1856+534~b with theoretical models. {\it Left panel}: best-fit model spectra for a Jupiter composition, cloud-free, atmosphere as a function of planet mass. {\it Right panel}: best-fit model spectra for a $0.7\,M_J$, Jupiter composition, atmosphere as a function of cloud-top pressure. The lightest model on the left panel is essentially the same as the lightest model on the right panel, since a 1000\,mbar cloud deck is too deep to influence the spectrum. All models assume a temperature of 163\,K, based on the equilibrium temperature of WD~1856+534~b \citep{Vanderburg2020}. The summary statistics do not include the data point near 0.55\,$\micron$ (see Section~\ref{subsec:comparison}). The best-fit models are essentially flat, requiring a mass similar to, or larger than, that of Jupiter ($M_p > 0.84\,M_J$ to 2\,$\sigma$) for clear models. Significant high-altitude cloud opacity can render lower masses compatible with the data (for $0.7\,M_J$, $P_{\mathrm{cloud}} <$ 120\,mbar to 2\,$\sigma$).}
\label{fig:model_spectra_comparison}
\end{figure*}

We computed a grid of time-averaged transmission spectra for model atmospheres spanning a range of planet masses and cloud-top pressures. We assume, for simplicity, an isothermal atmosphere at 163\,K with an atmospheric composition specified by mixing ratios (relative number densities) representative of Jupiter: $X_{\mathrm{CH_{4}}} = 1.8 \times 10^{-3}$, $X_{\mathrm{NH_{3}}} = 7 \times 10^{-4}$, and $X_{\mathrm{H_{2}O}} = 5 \times 10^{-4}$ \citep{Atreya2003}. We do not consider H$_2$O condensation, since even at these abundances the H$_2$O features are weak at optical wavelengths. Given that our observed spectrum is essentially flat (see Figure~\ref{fig:model_spectra_comparison}), we investigate the range of planet masses and/or cloud pressures that can render a Jupiter-like composition consistent with our observations. We considered planet masses from 0.5--5\,$M_J$ and cloud-top pressures from 1--1,000\,mbar. For each model, the reference planet radius at 10\,bar was the only parameter allowed to vary to obtain the best-fitting spectrum for each planet mass and cloud deck pressure.

\subsection{Atmospheric Inferences from WD~1856+534~b's Transmission Spectrum}

We compare representative model transmission spectra of WD~1856+534~b to our GMOS observations in Figure~\ref{fig:model_spectra_comparison}. Our observations have roughly the same uncertainty as the expected amplitude of several NH$_3$ features (e.g. 0.65, 0.73, and 0.79\,$\micron$) and a CH$_4$ feature (0.85\,$\micron$) for a Jupiter-mass clear atmosphere. However, we do not clearly detect any spectral features. Our flat spectrum can be explained by: (i) the planet mass is commensurate to, or larger than, that of Jupiter, thereby lowering the scale height through a higher surface gravity; or (ii) a high-altitude cloud deck is obscuring spectral features.

We explored the range of masses and cloud-top pressures consistent with our WD~1856+534~b transmission spectrum via simple frequentist model comparisons. For each model atmosphere, we fitted for the 10\,bar planetary radius (effectively a vertical offset in the model) for which the minimum $\chi^2$ was obtained. We convert the best-fitting $\chi^2$ into the equivalent p-value (using the incomplete gamma function) and derive how many standard deviations each model is `rejected' by the observations. These summary statistics are quoted in Figure~\ref{fig:model_spectra_comparison} for a series of clear atmospheres with different masses. We also show models for a $0.7\,M_J$ planet with different cloud-top pressures, showing how lower planetary masses can also fit the data if high-altitude clouds are present. We consider a 2\,$\sigma$ model rejection threshold for our quoted limits on the planet mass and cloud pressure.

Our analysis yields two key constraints on the atmosphere of WD~1856+534~b. First, we find a planet mass $> 0.84\,M_J$ (to 2\,$\sigma$) for clear atmospheres. Second, lower masses can still fit the data in the presence of a high-altitude cloud deck. We illustrate the latter point in Figure~\ref{fig:model_spectra_comparison} (right panel), showing that while a clear $0.7\,M_J$ model is rejected at 2.7\,$\sigma$, models with the same mass but with clouds can be compatible with our observed transmission spectrum ($P_{\mathrm{cloud}} <$ 120\,mbar for $0.7\,M_J$ to be rejected at less than 2\,$\sigma$). We note that these constraints do not include the 0.55\,$\micron$ data point, which appears discrepantly low compared to the rest of our transmission spectrum (see Figure~\ref{fig:model_spectra_comparison}). If this data point is included, we find that the lower mass bound is strongly dependent on this outlier ($> 1.9\,M_J$, to 2\,$\sigma$). We opt here for the more conservative mass bound consistent with the remainder of our spectrum, and discuss possible reasons for the low 0.55\,$\micron$ data point in Section~\ref{subsec:comparison}. 

Although our observed spectrum of WD~1856+534~b cannot distinguish between high-mass or high-altitude cloud scenarios, a comparison with Jupiter is informative. The highest cloud deck in Jupiter's atmosphere---formed from condensed NH$_3$---resides at $P_{\mathrm{cloud}}\sim$~700\,mbar \citep{Atreya2003}. At the slightly higher temperature of WD~1856+534~b, one would expect the NH$_3$ clouds to dissipate and the highest clouds to form from H$_2$O at $P_{\mathrm{cloud}} \sim$ 500\,mbar \citep[e.g.,][]{Burrows2004,MacDonald2018}. A similar picture exists for the cool brown dwarf WISE~0855, whose spectrum is suggestive of water ice clouds \citep{Morley2018}. For comparison, we require clouds at lower pressures than 120\,mbar to render masses as low as 0.7\,$M_J$ compatible with our observations. We suggest that a roughly Jovian, or super-Jovian, mass is more plausible than a sub-Jovian mass with a cloud deck at pressures $\sim 4 \times$ lower than expected from the condensation processes seen in Jupiter's atmosphere. Future observations at longer wavelengths, where molecular features are stronger, can provide a more conclusive means to measure the mass, cloud pressure, and atmospheric composition of WD~1856+534~b.

\section{Discussion \label{sec:dis}}

\subsection{Comparison to Previous Observations} \label{subsec:comparison}

\begin{figure}
\epsscale{1.2}
\plotone{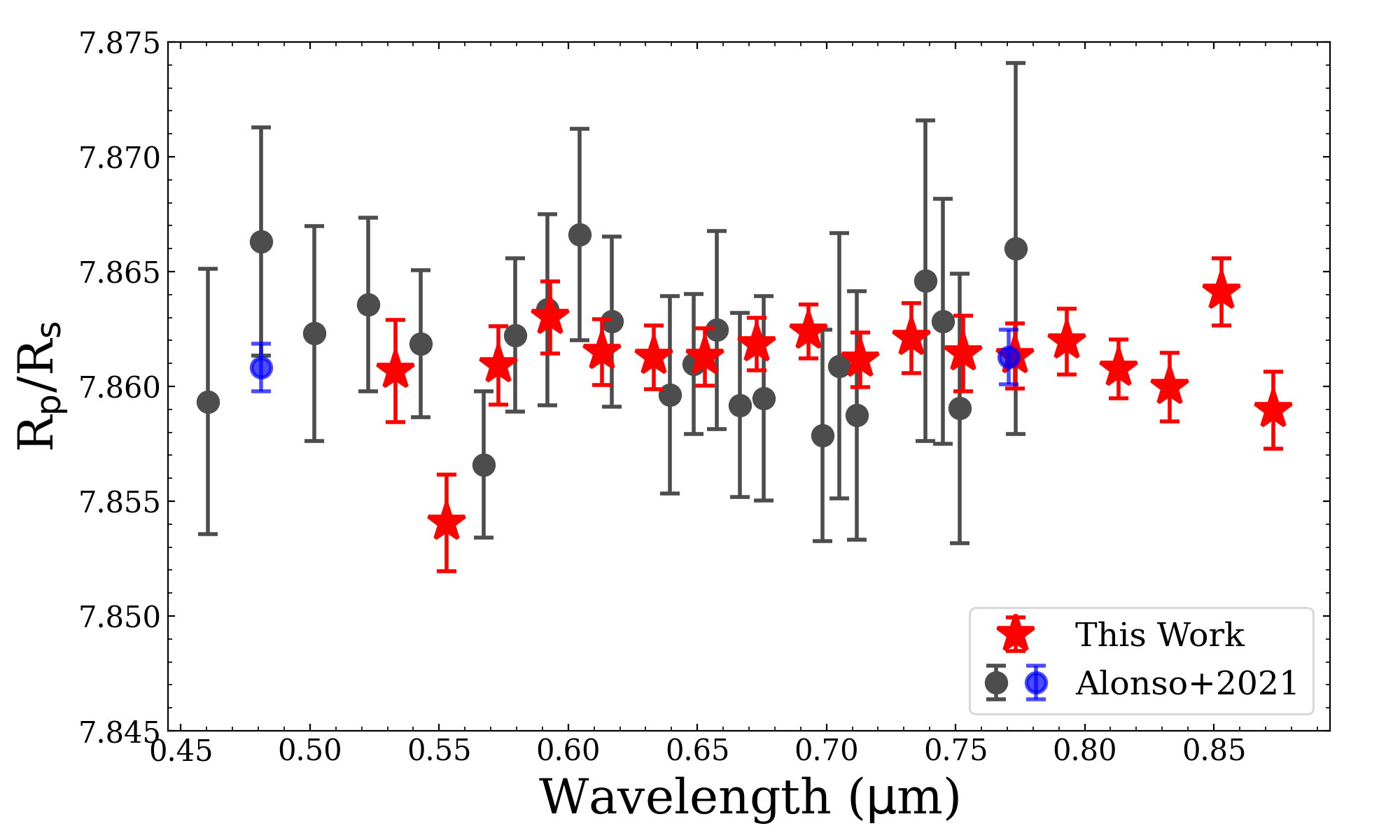}
\caption{Planet-to-star radius ratios (R$_\mathrm{p}$/R$_\mathrm{s}$) derived from this work compared to those in \citet{Alonso2021}. The black dots are transmission spectrum while the blue dots are photometry measurements. Both spectra are essentially flat, with a possible dip around 0.55~$\mathrm{\mu}$m.
}
\label{fig:comparison_Alonso2021}
\end{figure}

\citet{Alonso2021} presented a transmission spectrum of WD~1856+534~b with the GTC OSIRIS and EMIR instruments. Since their spectrum covers a similar wavelength range to our GMOS spectrum, we compare our respective spectra and the resulting atmospheric inferences. To put them on the same scale, the ratios in \citet{Alonso2021} are normalized to the medium value in this work. Figure~\ref{fig:comparison_Alonso2021} compares our transmission spectrum of WD~1856+534~b, expressed as $R_{\mathrm{p}, \lambda} / R_{\mathrm{s}}$, to the GTC spectrum from \citet{Alonso2021}. The morphology of our spectrum is in good agreement with \citet{Alonso2021}, especially the lack of a clear wavelength-dependence in both datasets. We see that our spectrum attains a higher signal-to-noise ratio (SNR)---especially at longer wavelengths ($\sim$ 3$\times$ higher for $\lambda > 6000$\,\AA)---due to our greater number of transits (8 vs.\ 2). Our observations also extend the transmission spectrum of WD~1856+534~b beyond 7800~\AA. 

\citet{Alonso2021} used the non-detection of a Rayleigh slope to place a lower bound on the mass of WD~1856+534~b, suggesting the planet is more massive than Jupiter (2.4\,M$_\mathrm{J}$ at 2$\sigma$ confidence). However, we find that significantly lower masses are allowed, even with our higher SNR spectrum: $M_{\mathrm{p}} > 0.84 M_\mathrm{J}$ at 2$\sigma$ confidence (for a clear atmosphere). There are two key reasons for this discrepancy, both associated with differences in how transmission spectra are modeled: (i) \citet{Alonso2021} assumed any optical wavelength dependence arises only from Rayleigh scattering, while we also include molecular absorption from a Jupiter-like atmosphere; and (ii) \citet{Alonso2021} used a formula from \citet{deWit2013} to relate the Rayleigh slope to planet mass, while we compute time-dependent radiative transfer accounting for the grazing transit. The aforementioned formula from \citet{deWit2013}
\begin{equation}
\alpha H = \frac{d R_{\rm p}(\lambda)}{d \ln \lambda}
\label{eq:deWit_slope}
\end{equation}
is derived assuming a planetary atmosphere fully overlapping a stellar disc during transit ($\alpha = -4$ for Rayleigh scattering). However, on average only $3.5\%$ of WD~1856+534~b's atmosphere overlaps its star a during transit (see Figure~\ref{fig:transit_animation}). Since the transmission spectrum slope is directly proportional to the fraction of the atmosphere overlapping the star, the mass corresponding to a given slope is lower than predicted by Equation~\ref{eq:deWit_slope} (and varies with time). Consequently, we suggest that the mass limit quoted by \citet{Alonso2021} is an overestimate. We also note that molecular absorption features can be stronger than Rayleigh scattering (see Figure~\ref{fig:model_spectra_comparison}), especially the NH$_3$ and CH$_4$ features at $\lambda > 7700$\,\AA. Mass constraints from optical transmission spectra should therefore consider both Rayleigh scattering and molecular absorption. In summary, accounting for the grazing transit geometry of WD~1856+534~b with a radiative transfer model is important for accurate mass constraints. With these considerations, we find that the mass of WD~1856+534~b is consistent with that of Jupiter (assuming a Jovian composition).

Intriguingly, both our spectrum and that of \citet{Alonso2021} show a discrepantly low data point near 5500\,\AA. In Figure~\ref{fig:comparison_Alonso2021}, we see that the planet-to-stellar radius ratio R$_\mathrm{p}$/R$_\mathrm{s}$ in our second GMOS bin (5430--5630\,\AA) is about 3$\sigma$ lower than the other spectroscopic bins. Similarly, the GTC transmission spectrum from \citet{Alonso2021} also has a dip around 0.567,$\mathrm{\mu}$m. This feature seems to be present in roughly half of our transits (see Figure~\ref{fig:rp/rwd}). If this 0.55\,$\mathrm{\mu}$m feature is physical and originates from WD~1856+534~b, such a negative transit depth could indicate an emission feature from the planetary nightside---potentially auroral emission, such as exhibited by Jupiter \citep{Yelle2004}. Future observations, especially at high spectral resolution, could probe wavelengths near 0.55\,$\mathrm{\mu}$m to examine if this feature is atmospheric in origin.

\begin{figure*}[ht!]
\plotone{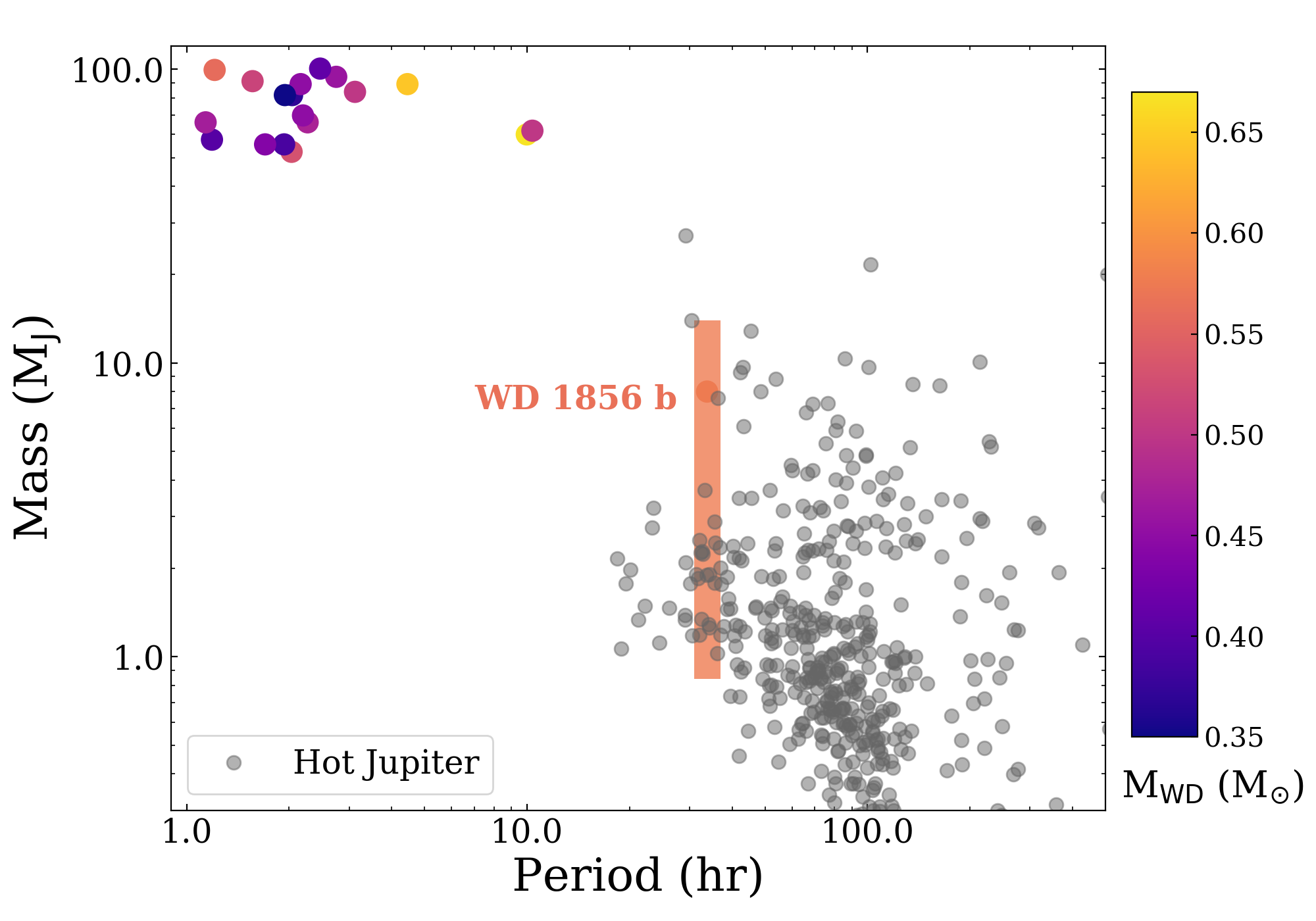}
\caption{Masses and periods for low-mass and short-period companions around white dwarfs (colored dots) and hot Jupiters around main sequence stars (grey dots). The white dwarf companions are color coded by the white dwarf masses, most of which are below 0.5~M$_\mathrm{\odot}$ as a result of the common envelope evolution. The error bars for the hot Jupiters are smaller than the sizes of the symbols. WD~1856+534~b sits close to the hot Jupiter population and the one closest to the nominal location of WD~1856+534~b is CoRoT-14~b.
}
\label{fig:MP}
\end{figure*}

\subsection{Formation Scenarios for WD~1856+534~b}

In this study, we constrained the mass of WD~1856+534~b to $>$ 0.84\,M$_\mathrm{J}$ (2\,$\sigma$ confidence, for clear atmospheres) and the mass of the white dwarf to 0.576~$\pm$~0.040~M$_\mathrm{\odot}$. To contextualize our results, we compiled a sample of white dwarfs with low-mass ($<$ 100 M$_\mathrm{J}$) and short-period ($<$ 100~hr) companions from the literature (\citealt{Nelson2018}, \citealt{FarihiChristopher2004}, \citealt{Parsons2015}, \citealt{Casewell2020} and \citealt{vanRoestel2021}). For the objects in \citet{Parsons2015}, the companion mass is estimated from the mass-spectral type relation of \citet{BaraffeChabrier1996}. We also include a sample of hot Jupiters around main sequence stars from the NASA Exoplanet Archive\footnote{\href{https://exoplanetarchive.ipac.caltech.edu/index.html/}{exoplanetarchive.ipac.caltech.edu/index.html}}. Figure~\ref{fig:MP} shows that there is some overlap between the hot Jupiter population and WD~1856+534~b. The mechanism behind how WD~1856+534~b arrived at its current 1.4~day orbit has been under intense study since its discovery. There are two general scenarios: (i) common envelope evolution, which is often invoked to explain the low-mass and short-period companions around white dwarfs; and (ii) dynamical evolution, which is a probable scenario to explain hot Jupiters. Our observations do not rule out either scenario, but here we discuss their pros and cons.

The common envelope scenario suggests that WD~1856+534~b formed relatively close to the white dwarf and survived the common envelope evolution by successfully expelling the envelope. However, the mass of WD~1856+534~b is small compared to other known post common envelope binaries and it might not have enough energy to expel the envelope---the main argument against the common envelope scenario offered by \citet{Vanderburg2020}. Recent studies find that WD~1856+534~b could have survived the common envelope evolution under a specific set of assumptions, or if there were multiple planets in the system all contributing to the envelope expulsion \citep{Lagos2021,Chamandy2021}. However, white dwarfs in these post common envelope binaries have a clear concentration below 0.55~M$_\mathrm{\odot}$, because their evolution has been truncated \citep{Rebassa-Mansergas2011}. Our updated white dwarf mass is 0.576~$\pm$~0.040~M$_\mathrm{\odot}$, which is close to the average mass of single white dwarfs. The total age (main sequence plus white dwarf cooling) of \object{WD~1856+534} is 8--10~Gyr, assuming single star evolution \citep{Lagos2021}. This new age estimate is also consistent with the results from the space motion analysis of the M dwarf pair \citep{Vanderburg2020}, suggesting that this system belongs to the thin disk of our Galaxy. 

Figure~\ref{fig:MP} shows a large gap, both in orbital periods and masses, between WD~1856+534~b and other low-mass and short-period companions around white dwarfs. This is partly an observational bias, as the transit detectability drops significantly for objects with larger orbital periods \citep{Gaudi2005}. However, another common way to find short-period companions around white dwarfs is via an infrared excess, which is not biased towards short-period objects and is limited by the sensitivity of infrared photometry \citep{Girven2011,Steele2011}. Indeed, five of the eighteen companions around white dwarfs shown in Figure~\ref{fig:MP} were first identified via their infrared excesses. Different all-sky surveys will help fill in the parameter space and return a more complete sample of low mass companions around white dwarfs.

The dynamical evolution scenario suggests that WD~1856+534~b arrived at its current location after the star evolved into a white dwarf. This is the favored scenario by \citet{Vanderburg2020}, as they identified several dynamical mechanisms to generate a small periastron distance and shrink the semi-major axis to bring WD~1856+534~b to its current location. Many previous studies have explored scattering mechanisms for bringing faraway planetesimals close enough to white dwarfs to produce their observed pollution \citep[e.g.][]{Veras2016}. Being part of a triple star system, the Kozai-Lidov mechanism is the preferred explanation for \object{WD~1856+534}. Simulations show that the Kozai-Lidov mechanism can perturb a planet originally located beyond 10~AU from the main sequence progenitor of \object{WD~1856+534} to the current location of WD~1856+534~b \citep{MunozPetrovich2020, OConnor2021,Stephan2020}. 

Figure~\ref{fig:MP} shows that WD~1856+534~b has some overlap with the parameter space occupied by the hot Jupiters, some of which likely also migrated due to the Kozai-Lidov effect \citep[e.g.][]{Petrovich2015,Anderson2016, DawsonJohnson2018}. The mass and orbital period of WD~1856+534~b match well with CoRoT-14~b, which is a 7.6~M$_\mathrm{J}$ planet on a 1.5~day orbit around an F9V star \citep{Tingley2011}. CoRoT-14~b has an eccentricity consistent with zero, but many other massive planets with short orbital periods have non-zero eccentricities, which is interpreted as a result of the long tidal circularization timescales \citep{Ferraz-Mello2015}. Similarly, one main uncertainty for dynamical evolution scenarios applied to WD~1856+534~b is whether tidal circularization can occur efficiently within the cooling age of the white dwarf \citep{VerasFuller2019}. However, if WD~1856+534~b ends up being more massive, it is less likely to be tidally disrupted during dynamical evolution, which is the main limiting factor for the formation of hot Jupiters via high-eccentricity migration \citep{Petrovich2015}. These considerations underscore the importance of obtaining an accurate measurement of WD~1856+534~b's mass in order to constrain its evolution history.

\subsection{Prospects for Further Characterization of WD~1856+534~b}

It is noteworthy that \object{WD~1856+534} is not polluted, even though 25--50\% white dwarfs display pollution attributed to planetary systems \citep{Zuckerman2003, Zuckerman2010,Koester2014a}. With our revised stellar parameters, the calcium abundance upper limit in \object{WD~1856+534}'s atmosphere is -11.2, which is low compared to other cool white dwarfs \citep{Hollands2018,Blouin2019}. The settling time in the atmosphere for calcium is about 1~Myr \citep{Dufour2017}, implying \object{WD~1856+534} has accreted very little, if any, material for the past 1~Myr. \object{WD~1856+534} also lacks an infrared excess from a dust disk, as shown in Figure~\ref{fig:SED}. Generally speaking, a white dwarf in a multiple star system or with a planetary system can frequently accrete materials due to gravitational interactions that shepherd small bodies to the white dwarf's tidal radius \citep[e.g.][]{Bonsor2011,Stephan2017}. For example, WD~1425+540 is in a wide binary system and has accreted material as far out as the Kuiper Belt \citep{Xu2017}. On the other hand, a massive, close-in planet such as WD~1856+534~b might act as a barrier to material that would otherwise pollute the white dwarf \citep{Veras2020}. Further study is needed to understand how WD~1856+534~b affects the evolution of minor bodies in the system. Looking forward, WD~1856+534~b's orbit will eventually shrink due to gravitational radiation and the system will become a cataclysmic variable in several trillion years \citep{PylyserSavonije1988}.

WD~1856+534~b offers the tantalizing prospect of characterizing a post main sequence giant planet. Despite the dim host \object{WD~1856+534} (G=17~mag), the large planet-star radius ratio can produce prominent absorption features in transmission spectra \citep[e.g.][]{loeb2013,Kozakis2020,Kaltenegger2020}. Though our optical spectrum does not reveal any absorption features, future infrared observations are promising. The host white dwarf is about 1 magnitude brighter in the near-infrared due to its cool temperature and molecules like NH$_3$ and CH$_4$ have stronger features in the near-infrared than at optical wavelengths \citep[e.g][]{Coles2019,Yurchenko2017}. Besides detecting features in transmission spectra, a direct detection of thermal flux from the planet could place an even stronger limit on the planetary mass. To investigate these possibilities WD~1856+534~b will be observed in the near-infrared by two programs in JWST Cycle 1\footnote{GO programs \href{https://www.stsci.edu/jwst/science-execution/program-information?id=2358}{\#2358} (PI: MacDonald), \href{https://www.stsci.edu/jwst/science-execution/program-information?id=2507}{\#2507} (PI: Vanderburg).}.

Detailed characterization of WD~1856+534~b's atmosphere will help to understand the origin of this enigmatic system. It will be interesting to see how close it resembles the cool giant planets in our solar system and cool brown dwarfs like WISE~0855. {\it TESS} is still ongoing and upcoming surveys, such as the Vera C. Rubin Observatory, will likely detect more transiting planets around white dwarfs \citep{Lund2018, CortesKipping2019}. With more objects filling in the empty parameter space in Figure~\ref{fig:MP}, we will have a better understanding about the properties of planets around white dwarfs.

\subsection{A New Method for Grazing Transit \\ Transmission Spectroscopy}

Motivated by the transit geometry of WD~1856+534~b, we introduced a new framework for transmission spectroscopy of planets with grazing transits ($b \gtrsim 1$). In Section~\ref{subsec:grazing_transit_spectra}, we proposed that the limb darkening corrected, time-averaged transmission spectrum (Figure~\ref{fig:grazing_transit_spectrum_extraction}) captures the time dependent nature of grazing transit spectra better than $(R_{\mathrm{p}, \lambda}/R_{\mathrm{s}})^2$. Time-averaged transmission spectra can be computed from radiative transfer models, such as the one introduced in Section~\ref{subsec:radiative_transfer}, accounting for the changing atmosphere area occulting a star (Figure~\ref{fig:transit_animation}). Our framework can be equally applied to other systems with grazing transits.

The only other exoplanet with a grazing transit and an observed transmission spectrum is WASP-67b \citep{Hellier2012,Bruno2018}. The transmission spectrum of WASP-67b shows muted H$_2$O features, which \citet{Bruno2018} attributed to cloud opacity or a metal-rich atmosphere. However, the atmospheric models used in that study assumed complete overlap of the planetary atmosphere with the star. We suggest that these muted H$_2$O features could instead be a consequence of the grazing transit geometry, as the amplitudes of spectral features change with time for grazing transits (see Figure~\ref{fig:transit_animation}). The method for grazing transit transmission spectra we introduce in this paper could be applied to systems such as WASP-67b in future studies.

Besides WASP-67b and WD~1856+534~b, some $\sim$ 20 confirmed exoplanets undergo grazing transits \citep{Davis2020}. Recently, {\it TESS} has detected several hot giant planets with grazing transits around bright stars, including TOI~216~b \citep{Kipping2019,Dawson2019}, TOI~1130~c \citep{Huang2020}, HIP 65Ab \citep{Nielsen2020}, and TOI~564~b \citep{Davis2020}. These planets are excellent targets for grazing transit spectroscopy. 

\section{Conclusion \label{sec:con}}

We performed optical transmission spectroscopy with Gemini/GMOS covering ten transits of WD~1856+534~b. Our main results are summarized as follows:

\begin{enumerate}
    \item Despite being a challenging system to observe (faint host star and short transit duration), we obtained a high quality optical transmission spectrum of WD~1856+534~b. We achieve a precision of 0.12\% in the phase-folded white light curve. Our results are generally consistent between different epochs. Besides cloud coverage, we find that the sky background plays a big role in the quality of the transmission spectrum.
    \item Due to the unique transit geometry of WD~1856+534~b, we show that the transmission spectrum itself is time-dependent. Therefore, we introduced the `limb darkening corrected, time-averaged transmission spectrum' as a potential alternative to the `traditional' ($R_{\mathrm{p}} / R_{\mathrm{s}}$)$^2$ spectrum. This method allows a direct comparison with atmospheric models and is applicable to all grazing systems. We also presented a modified radiative transfer prescription for modeling grazing transits.
    \item The spectrum of WD~1856+534~b shows no prominent spectral features, except a possible dip around 0.55~$\mathrm{\mu}$m. We find a planet mass $>$ 0.84\,$M_J$ (2$\sigma$), if WD~1856+534~b has a clear atmosphere with a Jovian composition, though clouds significantly higher than those on Jupiter can allow lower masses. Our mass limit is lower than the limit (2.4\,M$_\mathrm{J}$ at 2$\sigma$) found by \citet{Alonso2021}, even though our spectrum has higher quality. This can be explained by our different modeling approaches. Our mass limit accounts for the grazing transit geometry with a radiative transfer model, which is important to place accurate mass constraints.
    \item We confirm the H$\alpha$ detection in the white dwarf spectrum presented in \citet{Alonso2021}. We also revise the white dwarf parameters with the new infrared photometry. Now the age of \object{WD~1856+534} agrees with that of the co-moving M dwarf pair and this system likely belongs to the Galactic thin disk. 
    \item Our mass limit still leaves the origin of WD~1856+534~b as a mystery. The main challenge for the common envelope scenario is whether WD~1856+534~b could have successfully expelled the envelope; for the dynamical evolution scenario, it is unclear whether tidal circularization can occur efficiently without disrupting WD~1856+534~b. 
\end{enumerate}

With ongoing discoveries of more substellar objects around white dwarfs, further clues on the formation and evolution of post main sequence planetary systems will surely arise. The prospects for future characterization of WD~1856+534~b and its system are bright. For the tenacious world that survived the death of its star, our journey of discovery has only just begun.

\end{CJK}

\vspace{5mm}
{\it Acknowledgements.} We thank Roi Alonso for many helpful conversations and for sharing his team's GTC spectrum. We thank Greg Zeimann, Julia Scharw{\"a}chter, Kristin Chiboucas, Inger Jorgensen, Pier-Emmanuel Tremblay, Elena Cukanovaite, Neale Gibson, Jacob Bean, Jean-Michel Desert, Vatsal Panwar, Sarah Casewell and Erik Dennihy for useful discussions. HDL acknowledges support from the Villum Foundation. S.B. was supported by the Laboratory Directed Research and Development program of Los Alamos National Laboratory under project number 20190624PRD2

This research made use of \textsf{exoplanet} \citep{exoplanet} and its
dependencies \citep{exoplanet:agol20, exoplanet:arviz, exoplanet:astropy13,
exoplanet:astropy18, exoplanet:kipping13, exoplanet:luger18, exoplanet:pymc3,
exoplanet:theano}. This research has made use of the NASA Exoplanet Archive, which is operated by the California Institute of Technology, under contract with the National Aeronautics and Space Administration (NASA) under the Exoplanet Exploration Program.

This work is based on observations obtained at the international Gemini Observatory, a program of NSF's NOIRLab, which is managed by the Association of Universities for Research in Astronomy (AURA) under a cooperative agreement with the National Science Foundation on behalf of the Gemini Observatory partnership: the National Science Foundation (United States), National Research Council (Canada), Agencia Nacional de Investigaci\'{o}n y Desarrollo (Chile), Ministerio de Ciencia, Tecnolog\'{i}a e Innovaci\'{o}n (Argentina), Minist\'{e}rio da Ci{\^e}ncia, Tecnologia, Inova\c{c}\~{o}es e Comunica\c{c}\~{o}es (Brazil), and Korea Astronomy and Space Science Institute (Republic of Korea). The data is processed using the Gemini IRAF package. This work was enabled by observations made from the Gemini North telescope, located within the Maunakea Science Reserve and adjacent to the summit of Maunakea. We are grateful for the privilege of observing the Universe from a place that is unique in both its astronomical quality and its cultural significance.

\vspace{5mm}
\facilities{Gemini:Gillett (GMOS-N), Exoplanet Archive}

\software{Astropy \citep{Astropy2013,Astropy2018}, batman \citep{Kreidberg2015}, exoplanet \citep{exoplanet,exoplanet:kipping13,exoplanet:theano,exoplanet:pymc3,exoplanet:arviz,exoplanet:luger18,exoplanet:agol20}, POSEIDON \citep{MacDonald2017}, Scipy \citep{Scipy}, Matplotlib \citep{Matplotlib}, CMasher \citep{CMasher}}


\end{document}